\documentclass[11pt]{article}
\pdfoutput=1
\usepackage[centertags]{amsmath}
\usepackage[square, comma, sort&compress,numbers]{natbib}
\usepackage{array,multirow}
\numberwithin{equation}{section}
\usepackage{amssymb}
\usepackage{graphicx}
\usepackage{color}

\usepackage{epsfig}
\usepackage{wrapfig}
\usepackage{float}
\usepackage{tikz,pgf}
\usepackage{rotating}
\usetikzlibrary{shapes}
\usetikzlibrary{calc}
\usetikzlibrary{decorations.pathmorphing}
\usetikzlibrary{decorations.pathreplacing,shapes.misc}
\usetikzlibrary{positioning}
\usetikzlibrary{arrows}
\usetikzlibrary{decorations.markings}

\def\be{\begin{equation}}
\def\ee{\end{equation}}
\def\ba{\begin{array}}
\def\ea{\end{array}}
\def\bal{\begin{align}}
\def\eal{\end{align}}

\def\dps{\displaystyle}
\newcommand{\half}{\frac{1}{2}}

\def\eg{{\it e.g.}}

\def\a{\tilde{1}}

\def\1{\tilde{1}}
\def\2{\tilde{2}}
\def\3{\tilde{3}}

\newdimen\tableauside\tableauside=1.0ex
\newdimen\tableaurule\tableaurule=0.4pt
\newdimen\tableaustep
\def\phantomhrule#1{\hbox{\vbox to0pt{\hrule height\tableaurule
width#1\vss}}}
\def\phantomvrule#1{\vbox{\hbox to0pt{\vrule width\tableaurule
height#1\hss}}}
\def\sqr{\vbox{%
  \phantomhrule\tableaustep

\hbox{\phantomvrule\tableaustep\kern\tableaustep\phantomvrule\tableaustep}%
  \hbox{\vbox{\phantomhrule\tableauside}\kern-\tableaurule}}}
\def\squares#1{\hbox{\count0=#1\noindent\loop\sqr
  \advance\count0 by-1 \ifnum\count0>0\repeat}}
\def\tableau#1{\vcenter{\offinterlineskip
  \tableaustep=\tableauside\advance\tableaustep by-\tableaurule
  \kern\normallineskip\hbox
    {\kern\normallineskip\vbox
      {\gettableau#1 0 }%
     \kern\normallineskip\kern\tableaurule}%
  \kern\normallineskip\kern\tableaurule}}
\def\gettableau#1 {\ifnum#1=0\let\next=\null\else
  \squares{#1}\let\next=\gettableau\fi\next}

\newtheorem{prop}{Proposition}[section]
\newtheorem{lemma}[prop]{Lemma}
\newtheorem{definition}[prop]{Definition}

\newcommand{\bref}[1]{\textbf{\ref{#1}}}

\def\cD{\mathcal{D}}

\def\cF{\mathcal{F}}

\def\cO{\mathcal{O}}

\numberwithin{equation}{section} \makeatletter
\@addtoreset{equation}{section}

\def\ads{AdS_{3}}
\def\be{\begin{equation}}
\def\ee{\end{equation}}
\def\ba{\begin{array}}
\def\ea{\end{array}}

\def\dps{\displaystyle}

\def\ba{\begin{array}}
\def\ea{\end{array}}

\def\dps{\displaystyle}

\def\re{{\rm{Re}}}

\def\cl{c\to\infty}

\usepackage{jheppub}

\makeatletter
\def\@fpheader{\vspace{-.1cm}}
\makeatother

\title{Many-point classical conformal blocks  and \\ geodesic networks on the hyperbolic plane}

\author[a,b]{Konstantin\ Alkalaev\,}

\affiliation[a]{I.E. Tamm Department of Theoretical Physics, \\P.N. Lebedev Physical
Institute,\\ Leninsky ave. 53, 119991 Moscow, Russia}
\affiliation[b]{Moscow Institute of Physics and Technology, \\
Dolgoprudnyi, 141700 Moscow region, Russia}
\emailAdd{alkalaev@lpi.ru}

\abstract{We study  the semiclassical holographic correspondence between  $2d$ CFT $n$-point  conformal blocks and  massive particle configurations  in the asymptotically $\ads$ space. On the boundary we use the heavy-light approximation in which case two of primary operators are the  background for the other $(n-2)$ operators considered as fluctuations.  In the bulk the particle dynamics can be reduced to the hyperbolic time slice. Although lacking exact solutions we nevertheless  show that for any  $n$ the classical $n$-point conformal block is equal to the  length of the dual geodesic network connecting $n-3$ cubic vertices of worldline segments. To this end, both the bulk and boundary systems are reformulated as potential vector fields. Gradients of the conformal block and geodesic length are given respectively by accessory parameters of the monodromy problem and particle momenta of the on-shell worldline action  represented as a function of insertion points. We show that the accessory parameters and particle  momenta are constrained by two different  algebraic equation systems which nevertheless have the same roots thereby guaranteeing the correspondence.}

\preprint{FIAN-TD-2016-21}
\arxivnumber{}

\begin{document}

\maketitle
\flushbottom

\section{Introduction}

The AdS/CFT correspondence gives an effective prescription how to calculate CFT correlators from AdS action, at least in the saddle-point approximation \cite{Maldacena:1997re,Witten:1998qj,Gubser:1998bc}. It is interesting that the correspondence can be understood at a more structural level. CFT correlation functions can be decomposed into the theory-independent  conformal blocks completely fixed by conformal symmetry. It is natural to question what are the bulk counterparts of  conformal blocks. Recently, such dual objects were described in the case of $2d$ CFT conformal blocks considered in the limit of infinite conformal parameters $\Delta, \cl$ what corresponds to the semiclassical approximation on the gravity side. It was shown that on the Riemann sphere the limiting   conformal blocks called classical are equally described as lengths of particular geodesic networks stretched in the asymptotically $\ads$ space \cite{Hartman:2013mia,Fitzpatrick:2014vua,Asplund:2014coa,Caputa:2014eta,deBoer:2014sna,Hijano:2015rla,Fitzpatrick:2015zha,Perlmutter:2015iya,Alkalaev:2015wia,Hijano:2015qja,Alkalaev:2015lca,Banerjee:2016qca,Chen:2016dfb}. The essential ingredient here is the heavy-light approximation, where two of primary operators form  the background for the other operators \cite{Fitzpatrick:2014vua}. Depending on their conformal dimensions the background operators produce the angle deficit or BTZ black hole in the bulk so that perturbative operators correspond to  massive test particles.

The basic idea behind this kind of semiclassical AdS/CFT correspondence is quite simple. The semiclassical regime assumes that both the central charge and conformal dimensions tend to infinity such that ratios $\Delta/c$ are kept fixed. Then, the  conformal block $F(z|\Delta, c)$ is exponentiated $F(z|\Delta, c)\sim \exp\left({\frac{1}{c}f(z|\frac{\Delta}{c})}\right)$ to yield the classical conformal block $f(z|\frac{\Delta}{c})$ \cite{Zamolodchikov:1987ie}. On the gravity side,  the semiclassical path integral is dominated by classical paths describing geodesic motions of massive particles corresponding to primary operator insertions on the boundary \cite{Balasubramanian:1999zv,Louko:2000tp,Aref'eva:2016pew}.

The relation between classical conformal blocks and classical mechanics goes far beyond the AdS/CFT correspondence. Indeed, conformal blocks are known to be solutions of the Virasoro singular vector decoupling condition \cite{Belavin:1984vu}. On the other hand, the conformal blocks with arbitrary  conformal dimensions in the semiclassical limit were shown to satisfy the Painlev\'{e} VI equation which is just the decoupling condition represented in the Hamilton-Jacobi form  \cite{Litvinov:2013sxa}. It follows that constraining conformal dimensions within the heavy-light approximation the Painleve VI equation can be perturbatively reduced to the equation of motion for massive particles propagating in the asymptotically $\ads$ space.   

The study of the singular vector decoupling condition  brings to light many rich algebraic structures (see, \eg, \cite{Litvinov:2013sxa,Kashani-Poor:2013oza} and references therein). For example, the monodromy problem for the semiclassical decoupling condition expressed by the second order Fuchsian equation and the classical conformal blocks are deeply related \cite{Belavin:1984vu,Zamolodchikov:1987ie}. The Fuchsian equation $\psi^{''}(z) + T(z)\psi(z) = 0$ has the monodromy group fixed by the form of the stress-energy tensor $T(z)$, where conformal dimensions $\Delta$ are residues at the  second order poles, while the accessory parameters $c_i$ are residues at the simple poles. On the other hand, the accessory parameters are gradients  of the classical conformal block $c_i = \frac{\partial}{\partial z_i} f(z|\frac{\Delta}{c})$. It follows that  fixing the monodromy allows finding  the classical conformal block.             

Classical conformal blocks in the heavy-light approximation can be calculated within various approaches, \eg, using the monodromy method \cite{Hartman:2013mia,Fitzpatrick:2014vua,Hijano:2015rla,Alkalaev:2015lca,Banerjee:2016qca}, the Zamolodchikov recursion or the FKW global block method  \cite{Hijano:2015rla,Fitzpatrick:2015zha,Alkalaev:2015fbw}, or the AGT technique \cite{Alkalaev:2015wia}. One way or another, the resulting block reproduces the length of dual geodesic network. However, it seems that only the monodromy method conceptually explains why the correspondence holds, see, \eg, the discussion in \cite{Hijano:2015rla}. For example, the accessory parameter of the monodromy method can be interpreted as the momentum of a particle in the bulk whose worldline is attached  to the conformal boundary.\footnote{The accessory parameters had previously emerged in connection with mechanical momenta. See, \eg, \cite{Cantini:2002jw}, where the hamiltonian structure of the 2+1 gravity is discussed in the context of the so-called Polyakov conjecture and the Liouville theory.} This is quite natural because  the accessory parameter is the gradient of the classical conformal block, while external particle's momentum is the gradient of the on-shell worldline action. Exact relation based on the analysis of the accessory parameter/particle's momentum equations was given in \cite{Alkalaev:2015lca}.  
  
Our analysis in this paper focuses on proving the  AdS/CFT correspondence between classical conformal blocks and dual geodesic networks in the $n$-point case in the heavy-light approximation. 
We show that many-point blocks and dual lengths coincide up to logarithmic terms related to the conformal map from the complex plane to cylinder. It is important to note that we do not find  these functions explicitly. Instead, using the monodromy method on the boundary and the worldline approach  in the bulk we prove that the two descriptions of $n$-point configurations  are equivalent.

Intrinsically,  the proof of the equivalence is reduced to considering a potential vector field  $A_i(x) = \partial_i U(x)$, where $x_i$ are $n-2$  coordinates and  $U(x)$ is a potential. Potential vector field systems underlie both the bulk and boundary analysis. On the boundary, using the monodromy  method we identify the vector field components with the accessory parameters and the potential with the conformal block, while $x_i$ are insertion coordinates of $n-2$ perturbative primary operators. In the bulk, the network stretched in the angle deficit space has $n-2$ boundary attachment points $x_i$,  the vector field components are particular components of canonical momenta of massive test particles, while the potential is the on-shell worldline action identified with the geodesic length.

Both the accessory parameters and momenta are subjected to complicated algebraic equations arising respectively  as the monodromy conditions on the boundary and the least action principle in the bulk. Their exact solutions are known only in the 4-point case because the corresponding equation systems  are reduced to just a quadratic equation \cite{Fitzpatrick:2014vua,Hijano:2015rla}. In many-point case only approximate  solutions are available \cite{Hartman:2013mia,Alkalaev:2015wia,Alkalaev:2015lca,Banerjee:2016qca}.  In this paper we  prove that the two algebraic equation systems have the same roots. Therefore, the correspondence is guaranteed.

The paper is organized as follows. In Section \bref{sec:mon} we review the monodromy method for conformal blocks and fix our notation and conventions. In Section \bref{sec:pert} the monodromy problem is analyzed using the heavy-light perturbation theory.  In Section \bref{sec:multi} we  discuss the bulk space with the angle deficit and the dual geodesic network, see also  Appendix \bref{sec:B}. Here, we consider the Routh reduction of the  bulk dynamics to the hyperbolic time slice, we study a vertex of three lines on the hyperbolic plane, and analyze the minimal length condition and the so-called angular balance condition. The bulk/boundary correspondence in the $n$-point case is shown in Section \bref{sec:duality}. Proofs of lemmas and propositions formulated in Sections \bref{sec:net} and \bref{sec:duality} are collected in Appendices \bref{sec:prop1} -- \bref{sec:lemmaRE}. In the concluding Section \bref{sec:con}  we discuss our main results and some future directions.

\section{Classical blocks and the monodromy problem}
\label{sec:mon}

There are two basic ideas to compute $n$-point classical conformal blocks using the monodromy method. First, conformal blocks in a given channel are eigenfunctions of the monodromy operator associated with a particular contour on the punctured region. Second, in  the $\cl$ limit a particular degenerate operator of the quantum  $(n+1)$-point conformal block  effectively decouples yielding the classical $n$-point conformal block. Comparing the monodromy matrices along particular contours  computed before and after taking the semiclassical limit defines all gradients of the $n$-point classical conformal block in terms of coordinates of punctures. In this way, the problem    is reduced to solving first order differential equations (for review, see, \eg, \cite{Hartman:2013mia,Litvinov:2013sxa,deBoer:2014sna}).

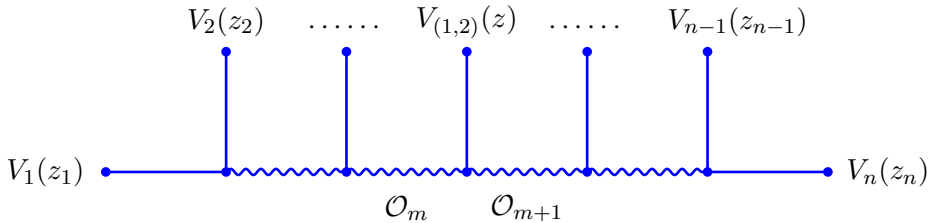
\begin{figure}[H]
\centering
\begin{tikzpicture}[scale=0.80]
\draw [blue, line width=1pt] (30,0) -- (32,0);
\draw [blue, line width=1pt] (32,0) -- (32,2);
\draw [blue, smooth, tension=1.0, line width=1pt, decorate, decoration = {snake, segment length = 2mm, amplitude=0.4mm}] (32,0) -- (34,0);
\draw [blue, line width=1pt] (34,0) -- (34,2);
\draw [blue, smooth, tension=1.0, line width=1pt, decorate, decoration = {snake, segment length = 2mm, amplitude=0.4mm}] (34,0) -- (36,0);
\draw [blue, line width=1pt] (36,0) -- (36,2);
\draw [blue, smooth, tension=1.0, line width=1pt, decorate, decoration = {snake, segment length = 2mm, amplitude=0.4mm}] (36,0) -- (38,0);
\draw [blue, line width=1pt] (38,0) -- (38,2);
\draw [blue, smooth, tension=1.0, line width=1pt, decorate, decoration = {snake, segment length = 2mm, amplitude=0.4mm}] (38,0) -- (40,0);
\draw [blue,line width=1pt] (40,0) -- (40,2);
\draw [blue,line width=1pt] (40,0) -- (42,0);


\draw (29,-0) node {$V_1(z_1)$};
\draw (32,2.5) node {$V_2(z_2)$};

\draw (34,2.4) node {$\cdots\cdots$};

\draw (36,2.5) node {$V_{(1,2)}(z)$};

\draw (38,2.4) node {$\cdots\cdots$};

\draw (35,-0.6) node {$\cO_m$};
\draw (37,-0.6) node {$\cO_{m+1}$};

\draw (43,-0) node {$V_n(z_n)$};
\draw (40.5,2.5) node {$V_{n-1}(z_{n-1})$};



\fill[blue] (32,0) circle (0.8mm);
\fill[blue] (30,0) circle (0.8mm);
\fill[blue] (32,2) circle (0.8mm);

\fill[blue] (34,0) circle (0.8mm);
\fill[blue] (34,2) circle (0.8mm);

\fill[blue] (36,0) circle (0.8mm);
\fill[blue] (36,2) circle (0.8mm);

\fill[blue] (38,0) circle (0.8mm);
\fill[blue] (38,2) circle (0.8mm);

\fill[blue] (40,0) circle (0.8mm);

\fill[blue]      (40,2) circle (0.8mm);

\fill[blue]       (42,0) circle (0.8mm);

\end{tikzpicture}
\caption{The $(n+1)$-point conformal block in a particular channel, where the degenerate operator fuses with $\cO_m$ to yield $\cO_{m+1}$ for any $m = 1,...\,, n-3$. There are $2(n-3)$ blocks of this type arising as solutions of the decoupling condition. In the limit $\cl$ the degenerate operator decouples and therefore the only surviving  block is given by that one shown  on Fig. \bref{block}. We fix $z_1=0$, $z_{n-1} = 1$, $z_n = \infty$. }
\label{n+1block}
\end{figure}

We consider the $(n+1)$-point correlation function $\langle V_{(1,2)}(y) V_{1}(z_1) \cdots V_n(z_n)\rangle$ on the Riemann sphere with one second level degenerate operator of  dimension  $\Delta_{(1,2)}$ in point $y$ and $n$ general primaries of  dimensions $\Delta_i$ in points $z_i$, $i = 1,...\,, n$. The correlation function satisfies the second order differential equation originating from the Virasoro singular vector decoupling condition   \cite{Belavin:1984vu}. The same is true for  conformal blocks because  the decoupling condition follows from Virasoro algebra only. The  space of solutions is two-dimensional, hence the monodromy operators are $2\times 2$ matrices. On the other hand, any $(n+1)$-point conformal block in a given channel has $n$ singular points and therefore there are $n$ independent monodromies.

We consider  the OPE associated with the set of concentric contours around a common center $z_1 = 0$. Inserting the degenerate primary $V_{(1,2)}(y)$ between primaries $V_{k}(z_k)$ and $V_{k+1}(z_{k+1})$ we fix a particular channel which means that $y$ should  lie on the contour enclosing insertion points $z_1, ...\,, z_m$:   
\be
\label{cons}
\text{Contour }\; \gamma_k\;\text{ encircles points }\;\; \{z_1,...\,, z_{k+1}\}\;, \qquad k = 1,...\,,n-3\;. 
\ee
Then,  points $z_{k+2}, ...\,, z_n$ are outside  contour $\gamma_k$ and, therefore, $\gamma_k \subset \gamma_{k+1}$. The resulting channels  are shown on Fig. \bref{n+1block}.

       
Remarkably, the OPE ties monodromy of solutions around particular contours  to dimensions of the exchanged operators in a particularly simple way. For the degenerate primary inserted as on Fig. \bref{n+1block} we find that the conformal block is  dominated by $(z_m - y)^{\tilde \Delta_{m+1} - \Delta_{(1,2)} - \tilde \Delta_{m}}$.
By the OPE argument, moving $y$ around $z_m$ is equivalent to moving around  insertion points of those operators which have been fused into the exchanged operator. Thus,  computing the monodromy of the above power-law function we easily find the monodromy along the contour $\gamma_k$ \eqref{cons}.
  
Indeed, using the Liouville parameterization\footnote{We change $(\Delta, c) \to (P, b)$ according to $\Delta(P) = \frac{c-1}{24} + P^2$ and $c = 1+ 6 (b+b^{-1})^2$ \cite{Belavin:1984vu}. The limit $\cl$ can equivalently be described as $b\to 0$.} we find that $\Delta_{(1,2)} = -1/2 -3b^2/4$, while conformal dimensions of exchanged operators are  related by  the fusion rule as $\tilde \Delta_{m+1}- \tilde \Delta_{m} = -b^2/4\, \pm\, ib P_{m}$  \cite{Belavin:1984vu}. Then, the monodromy matrix associated with $\gamma_k$ is given by  
\be
\label{M0}
\widetilde{\mathbb{M}}(\gamma_k) =  
\begin{pmatrix}
   e^{2\pi i M_{+k}} & 0 \\
   0 & e^{2\pi i M_{-k}} 
\end{pmatrix}\;,
\qquad 
M_{\pm k} = \half + \frac{b^2}{2}\pm ib P_{k-1}\;.
\ee

The classical conformal blocks arise in the limit when  the central charge and conformal dimensions simultaneously tend to infinity. Both external and exchanged dimensions $\Delta_m$ and $\tilde \Delta_n$ grow linearly with the charge $c$ in such a way that ratios $\epsilon_m = 6\Delta_m/c$ and $\tilde \epsilon_n = 6\tilde \Delta_n/c$ called classical dimensions remain fixed in $\cl$. Then, the quantum  conformal block is represented as an exponential of the classical conformal block \cite{Zamolodchikov:1987ie}. Operators with fixed classical dimensions are heavy, while those with vanishing classical dimensions are light.

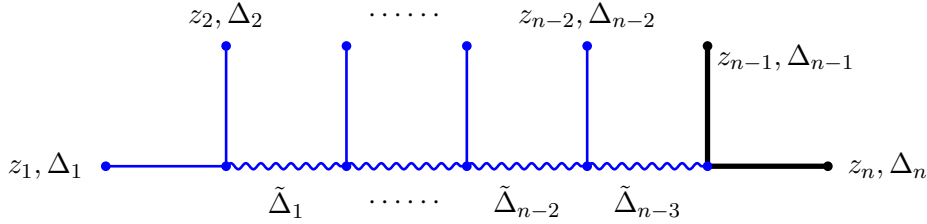
\begin{figure}[H]
\centering
\begin{tikzpicture}[scale=0.80]
\draw [blue, line width=1pt] (30,0) -- (32,0);
\draw [blue, line width=1pt] (32,0) -- (32,2);
\draw [blue, smooth, tension=1.0, line width=1pt, decorate, decoration = {snake, segment length = 2mm, amplitude=0.4mm}] (32,0) -- (34,0);
\draw [blue, line width=1pt] (34,0) -- (34,2);
\draw [blue, smooth, tension=1.0, line width=1pt, decorate, decoration = {snake, segment length = 2mm, amplitude=0.4mm}] (34,0) -- (36,0);
\draw [blue, line width=1pt] (36,0) -- (36,2);
\draw [blue, smooth, tension=1.0, line width=1pt, decorate, decoration = {snake, segment length = 2mm, amplitude=0.4mm}] (36,0) -- (38,0);
\draw [blue, line width=1pt] (38,0) -- (38,2);
\draw [blue, smooth, tension=1.0, line width=1pt, decorate, decoration = {snake, segment length = 2mm, amplitude=0.4mm}] (38,0) -- (40,0);
\draw [line width=2pt] (40,0) -- (40,2);
\draw [line width=2pt] (40,0) -- (42,0);


\draw (29,-0) node {$z_1, \Delta_1$};
\draw (32,2.5) node {$z_2, \Delta_{2}$};
\draw (38,2.5) node {$z_{n-2}, \Delta_{n-2}$};

\draw (35,2.5) node {$\cdots\cdots$};
\draw (43,-0) node {$z_n, \Delta_n$};
\draw (41.3,1.8) node {$z_{n-1}, \Delta_{n-1}$};

\draw (33,-0.6) node {$\tilde\Delta_1$};
\draw (39,-0.6) node {$\tilde\Delta_{n-3}$};
\draw (37,-0.6) node {$\tilde\Delta_{n-2}$};
\draw (35,-0.6) node {$\cdots\cdots$};


\fill[blue] (32,0) circle (0.8mm);
\fill[blue] (30,0) circle (0.8mm);
\fill[blue] (32,2) circle (0.8mm);

\fill[blue] (34,0) circle (0.8mm);
\fill[blue] (34,2) circle (0.8mm);

\fill[blue] (36,0) circle (0.8mm);
\fill[blue] (36,2) circle (0.8mm);

\fill[blue] (38,0) circle (0.8mm);
\fill[blue] (38,2) circle (0.8mm);

\fill[blue] (40,0) circle (0.8mm);

\fill       (40,2) circle (0.8mm);

\fill       (42,0) circle (0.8mm);

\end{tikzpicture}
\caption{The $n$-point conformal block. Two  bold black lines are  background heavy operators,  thin blue lines represent primary and exchanged perturbative heavy operators which are discussed  in Section \bref{sec:pert}.  }
\label{block}
\end{figure}

In our  case of the $(n+1)$-point conformal block all operators are supposed to be heavy while the degenerate operator is light,  $\lim_{b\to 0}\Delta_{(1,2)}= 1/2$. Thus, in the semiclassical limit it decouples from the other operators, while adjacent exchanged dimensions get equal $\lim_{b\to 0}(\tilde \Delta_m - \tilde\Delta_{m+1}) = 0$, see Fig. \bref{n+1block}. 
The limiting  $(n+1)$-point conformal block  factorizes as  
\be
\label{ccb5}
\cF(y,z| \Delta_m, \tilde \Delta_n) \,\Big |_{c\to\infty} \rightarrow \psi(y|z) \exp\big[-\frac{c}{6}f(z| \epsilon_i, \tilde \epsilon_j)\big]\;,
\ee
where we denoted $z = \{z_1,...\,,z_n\}$, function  $\psi(y|z)$ describes the semiclassical contribution of the degenerate operator, while   
the exponential factor $f(z| \epsilon_i, \tilde \epsilon_j)$ is the $n$-point classical conformal block which depends on external and exchanged classical conformal dimensions $\epsilon_i$ and $\tilde \epsilon_j$ \cite{Zamolodchikov:1987ie} (see also \cite{Zamolodchikov:1987ie,Hadasz:2005gk,Hartman:2013mia,Litvinov:2013sxa, deBoer:2014sna}).  We note that $(n+1)$-point conformal blocks can be considered in different channels arising  from different ways to insert the degenerate operator between other operators, see Fig. \bref{n+1block}. Such an ordering singles out a particular channel of the limiting $n$-point block \eqref{ccb5}, see Fig. \bref{block}.

Function $\psi(y|z)$ satisfies the Fuchsian equation arising from the decoupling condition
\be
\label{Tz}
\left[\frac{d^2}{dy^2}  + T(y|z)\right]\psi(y|z) = 0\;,\qquad \text{where} \qquad T(y|z) = \sum_{i=1}^n \frac{\epsilon_i}{(y-z_i)^2} +  \sum_{i=1}^n \frac{c_i}{y-z_i}\;.
\ee
The accessory parameters $c_{i}$ are gradients of the classical $n$-point conformal block 
\be
\label{acs}
c_i (z) = \frac{\partial f(z)}{\partial z_i}\;,\qquad i = 1,...\,,n\;,
\ee
and satisfy the linear  constraints
\be
\label{threecons}
\sum^n_{i = 1} c_i = 0\;,
\qquad
\sum^n_{i = 1} (c_i z_i + \epsilon_i) = 0\;,
\qquad
\sum^n_{i = 1} (c_i z_i^2 + 2\epsilon_i z_i) = 0\;.
\ee
It follows that expanding the stress-energy tensor around $y = \infty $ we find no terms $1/y^{l}$, $l = 1,2,3$ and therefore  near the infinity $T(y)\sim 1/y^4$. Choosing $c_2,...\,, c_{n-2}$ as independent parameters and fixing, accordingly,   $z_1 = 0, z_{n-1} = 1, z_n = \infty$, the above constraints are solved as \cite{Alkalaev:2015lca,Banerjee:2016qca}
\begin{align}
& c_1 = -\sum^{n-2}_{i=2}\big[c_i(1-z_i)- \epsilon_i\big] +\epsilon_1+\epsilon_{n-1}- \epsilon_{n}  \;,\label{ccc}\\
&c_{n-1} = - \sum_{i=2}^{n-2}c_i z_i + \epsilon_{n} -\sum_{i=1}^{n-1} \epsilon_i \;,
\qquad
c_n = 0\label{ccc1}\;.
\end{align}
Then, the stress-energy tensor $T(y|z)$ takes the form 
\be
\label{Tzred}
T(y|z) = \sum_{i=1}^{n-1}\frac{\epsilon_i}{(y-z_i)^2}+\sum_{i=2}^{n-2}\,c_i\, \frac{ z_i (z_i - 1)}{y(y - z_i) (y - 1)} + \frac{ \epsilon_n- \sum_{i=1}^{n-1} \epsilon_i}{y (y - 1)}\;.
\ee
The corresponding  Fuchsian equation still  has $n$ regular singular points around which we compute the monodromy. Continuing solutions $\psi(y|z)$ along the contours $\gamma_k$ \eqref{cons} on the punctured $y$-plane and comparing the resulting monodromy matrices with  \eqref{M0} we can find the accessory parameters as functions of  classical conformal dimensions and insertion points.

\section{Heavy-light perturbation theory} 
\label{sec:pert} 
 
Though the general solution to the Fuchsian equation with $n$ singular points is unknown we can try to use perturbation theory. The idea is to consider $k$ heavy operators as the background for the rest $n-k$ heavy operators. It follows that the corresponding dimensions are constrained as $\epsilon_{pert}/\epsilon_{back}\ll 1$, where $\epsilon_{back}$ are dimensions of the background operators, $\epsilon_{pert}$ are dimensions of the perturbative operators.  Obviously, the simplest cases are given by $k=2$ or $k=3$ background operators. The case of $k=1$ background operator is trivial because one-point functions on the sphere are vanishing. In the $k=2$ case, the background conformal dimensions should be equal to each other for the corresponding  two-point function to be non-vanishing. In order to apply perturbation theory in the $k\geq 4$ case we have to know exact solutions of the Fuchsian equation with $k$ singularities.    

Following \cite{Fitzpatrick:2014vua,Hijano:2015rla,Alkalaev:2015lca,Banerjee:2016qca} we consider the case of $k=2$ background operators (bold black lines on Fig. \bref{block}). Let  $ \epsilon_{n-1} = \epsilon_{n}\equiv \epsilon_h$ be the background heavy dimension, while $\epsilon_i$, $i=1,...\,,n-2$ be perturbative heavy dimensions. It is assumed that $\epsilon_i/\epsilon_h \ll 1$. Then, the Fuchsian equation \eqref{Tz} with the stress-energy tensor \eqref{Tzred} can be solved perturbatively. We expand all  functions as
\begin{align}
&\psi(y,z) = \psi^{(0)}(y,z)+ \psi^{(1)}(y,z) +\psi^{(2)}(y,z)+ ... \;,\label{expan0}\\
&T(y,z) = T^{(0)}(y,z)+ T^{(1)}(y,z)+T^{(2)}(y,z)+ ... \;,\label{expan1}\\
&c_i(z) = c_i^{(0)}(z)+ c_i^{(1)}(z) +c_i^{(2)}(z)+ ... \;, \label{expan2}
\end{align}
where  expansion parameters are perturbative heavy  dimensions. The accessory parameter expansion starts with terms linear in the conformal dimensions so that $c_i^{(0)}= 0$. For the sake of simplicity, from now on we denote $c_i^{(1)}(z) := c_i(z)$. The  classical conformal block \eqref{ccb5} is similarly expanded, 
\be
\label{pertblock}
f(z) = f^{(0)}(z)+ f^{(1)}(z)+f^{(2)}(z)+ ...\;,
\ee
in the way consistent with  expansion \eqref{expan2} and relation \eqref{acs}. The zeroth approximation corresponds to the classical conformal block of 2-point functions of the background operators. Hence,  $f^{(0)}(z) = 0$ and the first non-trivial correction is given by $f^{(1)}(z)$. By analogy with the first order accessory parameters we  denote $f^{(1)}(z) : = f(z)$ and therefore the relation \eqref{acs} remains unchanged.

\subsection{Solving the Fuchsian equation} 

Using \eqref{expan0} -- \eqref{expan2} we find that the perturbatively expanded  Fuchsian equation yields a chain of inhomogeneous linear equations $\cD_0 \psi^{(s)}(y,z) + T^{(s)} \psi^{(s-1)}(y,z) = 0$, $s=0,1,2,...$,   with a unique differential part given by the operator $\cD_0 = d^2/dy^2   + T^{(0)}(y)$. The lowest order equations are given by
\be
\label{deceq}
\ba{l}
\dps
\cD_0 \psi^{(0)}(y,z) = 0\;,
\qquad 
\cD_0 \psi^{(1)}(y,z) + T^{(1)} \psi^{(0)}(y,z) = 0 \;,
\ea
\ee
where the stress-energy components are read off from \eqref{Tzred}
\be
\label{T0}
T^{(0)}(y) = \frac{\epsilon_h}{(y-1)^2}\;,
\qquad
T^{(1)}(y,z) = \sum_{i=1}^{n-2}\frac{\epsilon_i}{(y-z_i)^2}+\sum_{i=2}^{n-2}c_i \frac{ z_i (z_i - 1)}{y(y - z_i) (y - 1)} - \frac{ \sum_{i=1}^{n-2} \epsilon_i}{y (y - 1)}\;,
\ee
with the convention  $z_1 = 0$. There are two zeroth order branches $\psi_{\pm}^{(0)}(y,z) = (1-y)^{(1\pm\alpha)/2}$, where
\be
\label{alpha}
\alpha  = \sqrt{1-4 \epsilon_h}\;.
\ee
Using the zeroth order solution we find in the first order that  
\be
\label{fos}
\psi^{(1)}_{\pm}(y,z) =  \frac{1}{\alpha}\psi_{+}^{(0)}(y) \int dy\, \psi^{(0)}_-(y)T^{(1)}(y,z)\psi_{\pm}^{(0)}(y)   - \frac{1}{\alpha} \psi_{-}^{(0)}(y) \int dy\, \psi^{(0)}_+(z)T^{(1)}(y,z)\psi_{\pm}^{(0)}(y)\;.  
\ee
Corrections $\psi_{\pm}^{(1)}$ have  branch points inherited from those of $\psi_{\pm}^{(0)}(y)$ and  $T(y,z)$. 

\subsection{Computing monodromies} 

Now, we  continue the perturbative solution  $\psi = \psi^{(0)}+ \psi^{(1)}+...$
along particular contours \eqref{cons}. Encircling  the branch points of the solution we define the monodromy  as  $\mathbb{M}(\gamma): \psi \to \mathbb{M}(\gamma)\psi$, where the monodromy matrix is also expanded as $\mathbb{M} = \mathbb{M}_0 + \mathbb{M}_1 + \ldots\;$. In components,  
\be
\gamma_k\;:\qquad \begin{pmatrix}
   \psi_{+} \\
   \psi_{-} 
\end{pmatrix}
\rightarrow 
\begin{pmatrix}
   M_{++}(\gamma_k) & M_{+-}(\gamma_k) \\
   M_{-+}(\gamma_k) & M_{--}(\gamma_k) 
\end{pmatrix}
\begin{pmatrix}
   \psi_{+} \\
   \psi_{-} 
\end{pmatrix}\;\;.
\ee
Expanding both the solution and the monodromy matrix as above we find that the zeroth order matrix  $\mathbb{M}_0$ defines the monodromy of $\psi^{(0)}(y)$ with the branch point $y=1$. However, the contours $\gamma_k$ \eqref{cons} do not enclose this point, and, therefore, $\mathbb{M}_0 = \mathbb{I}$. It follows that the  perturbative solution represented as a linear combination of the zeroth order solutions with the integral coefficients \eqref{fos} fits well the monodromy computation. Indeed, the whole computation is reduced to evaluating the integrals  along the contours $\gamma_k$,  
\begin{align}
&I^{(k)}_{\pm+}(z) = +\frac{1}{\alpha}\int_{\gamma_{k}} dy\; \psi^{(0)}_{-}(y)T^{(1)}(y,z)\psi_{\pm}^{(0)}(y)\;,\\
&I^{(k)}_{\pm-}(z) = -\frac{1}{\alpha}\int_{\gamma_{k}} dy\; \psi^{(0)}_{\pm}(y)T^{(1)}(y,z)\psi_{-}^{(0)}(y)\;.
\end{align}
Substituting the stress-energy tensor correction $T^{(1)}(y,z)$ given by \eqref{T0} and using the residue theorem we have 
\begin{align}
&I^{(k)}_{+-}  =\frac{2\pi i}{\alpha}\Big[\alpha \epsilon_1+\sum_{i=2}^{n-2}(c_i(1-z_i)-\epsilon_i)-\sum_{i=2}^{k+1}(1-z_i)^\alpha(c_i(1-z_i)-\epsilon_i(1+\alpha))\Big]\;,\label{I+-}
\\
&I^{(k)}_{++} = \frac{2\pi i}{\alpha}\sum_{i=k+2}^{n-2}\big[c_i (1-z_i)-\epsilon_i\big]\;,
\qquad
I^{(k)}_{-+}  = I^{(k)}_{+-}\big|_{\alpha \rightarrow -\alpha}\;,
\quad
I^{(k)}_{--} =  I^{(k)}_{++}\big|_{\alpha \rightarrow -\alpha}\;. \label{I++}
\end{align}
For a given number of insertion points $n$  integrals $I_{++}$ and $I_{--}$  over the maximal contour $\gamma_{n-3}$ are always zero, $I^{(n-3)}_{++} =- I^{(n-3)}_{--} =0$. It follows that elements of the first order correction  $\mathbb{M}_1$ are just the contour integrals, $M_{\pm\pm}(\gamma_k) = I_{\pm\pm}^{(k)}$. The second order monodromy matrix $\mathbb{M} = \mathbb{M}_0+\mathbb{M}_1$ is therefore given by
\be
\label{M}
\mathbb{M}(\gamma_k) = \begin{pmatrix}
   1+ I^{(k)}_{++} & I^{(k)}_{+-} \\
   I^{(k)}_{-+} & 1+I^{(k)}_{--} 
\end{pmatrix}\;.
\ee

\subsection{Eigenvalue condition}
\label{sec:eigen}
Matrices $\mathbb{M}(\gamma_k)$ \eqref{M} on the one hand and matrices $\widetilde{\mathbb{M}}(\gamma_k)$ \eqref{M0} at infinite central charge and small perturbative classical dimensions on the other hand  describe the same monodromy associated to continuation of the degenerate operator along the contours $\gamma_k$. Equating the corresponding eigenvalues we arrive at the system of equations on conformal dimensions of exchanged operators, insertion points, and accessory parameters.  

We consider the monodromy \eqref{M0} semiclassically. The limiting matrix reads  
\be
\label{M111}
\lim_{b\to 0}\widetilde{\mathbb{M}}(\gamma_k) =  
\begin{pmatrix}
   e^{2\pi i M_{-k}} & 0 \\
   0 & e^{2\pi i M_{+k}} 
\end{pmatrix}\;,
\qquad 
M_{\pm k} = \half \pm \half \sqrt{1-4\tilde \epsilon_{k}} \;.
\ee
Within the perturbation theory, the eigenvalues of \eqref{M111} up to linear order in dimensions of the exchanged operators  are given by $\lambda_{\pm k} = 1\, \pm\, 2\pi i\, \tilde\epsilon_{k}$. To diagonalize matrices \eqref{M} we solve the characteristic equation $\det\left(\mathbb{M}(\gamma_k) - \lambda_k  \mathbb{I}\right) = 0$. Using  \eqref{I++} we find that the eigenvalues are defined by the quadratic equation  $(1-\lambda_k)^2 = I^{(k)}_{++}+ I^{(k)}_{+-}I^{(k)}_{-+}$. Equating two sets of eigenvalues we arrive at the following  system
\be
\label{acc_sys}
\left(I_{++}^{(k)}\right)^2 + I_{-+}^{(k)}\,I_{+-}^{(k)} =   - 4\pi^2 \tilde\epsilon^2_{k}\;,
\qquad k = 1,...\,, n-3\;.
\ee

In what follows,  equations  \eqref{acc_sys} supplemented by equations \eqref{ccc} are referred to as the accessory equations.  Recalling the form of the contour integrals \eqref{I+-}--\eqref{I++} we find out that \eqref{acc_sys} is the system of quadratic equations  with coefficients depending on punctures $z_k$ and  conformal dimensions $\alpha, \epsilon_i, \tilde \epsilon_j$. Solving them we can unambiguously express the accessory parameters as functions of conformal  dimensions and coordinates of the punctures. We discuss the accessory parameters equations in Section \bref{sec:accsys} and their dual interpretation in Section \bref{sec:duality}. Here we stress  that classical conformal dimensions of external and exchanged operators in \eqref{acc_sys} are arbitrary. However, we can impose various constraints to facilitate solving the accessory parameters  equations using  approximation techniques \cite{Fitzpatrick:2014vua,Hijano:2015rla,Alkalaev:2015lca,Alkalaev:2015fbw,Banerjee:2016qca}.

\section{Geodesic networks on the hyperbolic plane}
\label{sec:multi}

Perturbative conformal blocks can be represented in terms of massive point particles propagating in the three-dimensional space described by the metric  \cite{Fitzpatrick:2014vua}
\be
\label{metric}
ds^2  = \frac{\alpha^2}{\cos^2 \rho}\Big( dt^2 +\sin^2\rho d\phi^2 +\frac{1}{\alpha^2} d\rho^2 \Big)\;,
\ee
where  $t\in \mathbb{R}$, $\rho \in [0,\pi/2]$, $\phi \in [0,2\pi)$. The space contains a conical singularity measured by the angle deficit  $2\pi(1-\alpha)$, where $\alpha \in (0,1]$. 
The metric \eqref{metric} describes the constant negative curvature space with the topology $\mathbb{R}\times {\cal C}_2$.  Two-dimensional slices $\phi = const$ and $t = const$  are also negative curvature spaces identified with the punctured hyperbolic plane, where the puncture corresponds to the conical singularity projected on two dimensions. The  conformal boundary reached at $\rho \rightarrow \pi/2$ is the Euclidean cylinder with local coordinates $t$ and $\phi$.  The presence of  conical singularity breaks the global $\ads$  isometry ($\alpha=1$) down to Abelian isometry  $\mathbb{R} \oplus o(2)$ generated by two Killing vectors $\partial_t$ and $\partial_\phi$. On the conformal boundary, the Abelian isometry is enhanced to full Virasoro algebra.

\begin{figure}[H]
\centering
\begin{tikzpicture}[line width=1pt,scale=0.80]
\draw (0,0) circle (3cm);

\foreach \a in {1,2,...,40}{
\draw (\a*360/40: 3.5cm) coordinate(N\a){};
\draw (\a*360/40: 3.7cm) coordinate(G\a){};
\draw (\a*360/40+5: 3.5cm) coordinate(D\a){};

\draw (\a*360/40: 1.3cm) coordinate(H\a){};
\draw (\a*360/40: 1.35cm) coordinate(Q\a){};

\draw (\a*360/40: 4cm) coordinate(A\a){};
\draw (\a*360/40: 3cm) coordinate(K\a){};
\draw (\a*360/40: 2.5cm) coordinate(F\a){};
\draw (\a*360/40: 2cm) coordinate(L\a){};
\draw (\a*360/40: 1.5cm) coordinate(I\a){};
\draw (\a*360/40: 1.1cm) coordinate(J\a){};
\draw (\a*360/40: 1cm) coordinate(M\a){};
\draw (\a*360/40: 0.8cm) coordinate(C\a){};
;}


\draw[blue] plot [smooth, tension=1.0, line width=1pt] coordinates {(K35) (L34) (I30)};
\draw[blue] plot [smooth, tension=1.0, line width=1pt] coordinates {(K27) (L28) (I30)};
\draw[blue] plot [smooth, tension=1.0, line width=1pt] coordinates {(K23) (L23) (M25)};
\draw[blue] plot [smooth, tension=1.0, line width=1pt] coordinates {(K20) (L19) (M19)};
\draw[blue] plot [smooth, tension=1.0, line width=1pt] coordinates {(K16) (L14) (M12)};


\draw [blue, smooth, tension=1.0, line width=1pt, decorate, decoration = {snake, segment length = 2mm, amplitude=0.4mm}] (M12)  -- (0,0);

\draw  [blue, smooth, tension=1.0, line width=1pt, decorate, decoration = {snake, segment length = 2mm, amplitude=0.4mm}]  (I30) -- (J29) -- (M25);

\draw  [blue, smooth, tension=1.0, line width=1pt, decorate, decoration = {snake, segment length = 2mm, amplitude=0.4mm}]  (M12) -- (J16) -- (M19);

\draw  [blue, smooth, tension=1.0, line width=1pt, decorate, decoration = {snake, segment length = 2mm, amplitude=0.4mm}]  (J19) -- (M25);

\fill[blue] (I30) circle (0.8mm);
\fill[blue] (M25) circle (0.8mm);
\fill[blue] (K35) circle (0.8mm);
\fill[blue] (K27) circle (0.8mm);
\fill[blue] (K23) circle (0.8mm);
\fill[blue] (0,0) circle (0.8mm);
\fill[blue] (M19) circle (0.8mm);
\fill[blue] (M12) circle (0.8mm);
\fill[blue] (K20) circle (0.8mm);
\fill[blue] (K16) circle (0.8mm);

\draw (-3.4,-1.6) node {$w_3,  \epsilon_3$};
\draw (N27) node {$w_2,  \epsilon_2$};
\draw (G35) node {$w_1,  \epsilon_1$};
\draw (A16) node {$w_{n-2},  \epsilon_{n-2}$};

\draw (0.5,0.4) node {$\tilde\epsilon_{n-3}$};
\draw (-0.6,-1.2) node {$\tilde\epsilon_{1}$};
\draw (-1.3,-0.2) node {$\tilde\epsilon_{2}$};


\draw (D21) node {.};
\draw (N21) node {.};
\draw (D20) node {.};
\draw (N20) node {.};
\draw (D19) node {.};
\draw (N19) node {.};
\draw (D18) node {.};
\draw (N18) node {.};
\draw (D17) node {.};
\draw (N17) node {.};
\draw (D16) node {.};

\draw (H17) node {.};
\draw (Q16) node {.};
\draw (H15) node {.};

%
%
%

\fill[blue] (I30) circle (0.8mm);
\fill[blue] (M25) circle (0.8mm);
\fill[blue] (K35) circle (0.8mm);
\fill[blue] (K27) circle (0.8mm);
\fill[blue] (K23) circle (0.8mm);
\fill[blue] (0,0) circle (0.8mm);
\fill[blue] (M19) circle (0.8mm);
\fill[blue] (M12) circle (0.8mm);
\fill[blue] (K20) circle (0.8mm);
\fill[blue] (K16) circle (0.8mm);

\end{tikzpicture}
\caption{Network of geodesic lines on the hyperbolic disk. Solid and wave lines denote respectively external ($\epsilon_m$) and exchanged ($\tilde \epsilon_k$) particles, dotted lines denote the middle part of the graph. The boundary attachment points are $w_m$, $m=1,...\,,n-2$.} 
\label{bulk}
\end{figure}
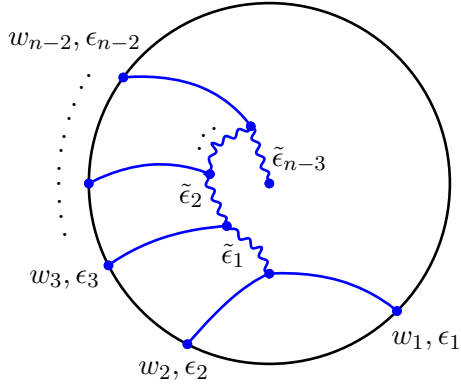

A massive particle on the angle deficit space with the interval \eqref{metric} is described by the worldline action   
$S = \dps\epsilon\int d\lambda \, \sqrt{g_{tt} \dot{t}^2+g_{\phi\phi} \dot{\phi}^2+g_{\rho\rho} \dot{\rho}^2}$, where $\epsilon$ is a classical conformal dimension identified with a mass, the metric coefficients are read off from \eqref{metric}, $\lambda$ is the  evolution parameter and  the dot  denotes differentiation with respect to $\lambda$, see Appendix \bref{sec:B} for more details. The Abelian isometry guarantees that coordinates $\phi$ and $t$ are cyclic, {\it i.e.} $\delta  S/\delta  \phi \equiv 0$ and $\delta S/\delta t \equiv  0$. It follows that the original mechanics can be reduced to a simpler system described by the Routhian function, which means that we have to perform a partial Legendre transformation with respect to $\dot\phi$ and $\dot t$. Choosing the partial constraint $\dot t = 0$ we arrive at the Routhian action, which describes a massive particle moving on the punctured hyperbolic disk, 
\be
\label{disk}
S = \int d \lambda\, L\;, \qquad L =\epsilon\, \sqrt{\alpha^2 \tan^2 \rho\, \dot{\phi}^2+\sec^2 \rho \,\dot{\rho}^2}\;.
\ee
The residual isometry is given by $sl(2, \mathbb{R})$ at $\alpha=1$ and  $o(2)$ at $\alpha\neq 1$.

We consider a set of massive point particles propagating around the conical singularity. They interact to each other forming cubic vertices of worldlines. Of course, there are other possible types of interaction including quartic and higher vertices. However, the block/length correspondence singles out only cubic vertices. There are $2n-5$  particles  corresponding to the total number of external/exchanged lines of the dual $n$-point conformal block diagram shown on Fig. \bref{block}. External $n-2$ worldlines are attached to the conformal boundary at fixed points $w=(w_1,...\,,w_{n-2})$, where $w = \phi+i t$. Exchanged $n-3$ worldlines are stretched between vertices except for the radial line which ends at the center of the disk. The resulting geodesic network on the hyperbolic disk is shown on Fig. \bref{bulk} \cite{Alkalaev:2015wia}. It can be obtained by copy-pasting the conformal block diagram on Fig. \bref{block} into the disk so that two background operator lines shrink to the point identified with the center of the disk, while the insertion points of conformal primaries go to the boundary attachment points.\footnote{The geodesic network on Fig. \bref{bulk} can be obtained through the geodesic Witten diagram studied in \cite{Hijano:2015qja,Nishida:2016vds}.}

%

Geodesics on the hyperbolic time slice \eqref{disk} are most easily described using the Poincare disk model. In this case, these are segments of circles perpendicular to the boundary,  including circles of infinite radius (the radial line on Fig. \bref{bulk}). Geodesic lengths are explicitly known as functions of endpoints, see, \eg, formula \eqref{lambda}.

\subsection{Cubic vertex and triangle inequalities} 
\label{sec:vertex}

\begin{figure}[H]
\centering
\begin{tikzpicture}[line width=1pt,scale=0.80]

\draw [->] (0,0) --(0,1);
\draw      (0,1) --(0,2);

\draw      (0,0) -- (-0.6,-0.6);
\draw [<-] (-0.6,-0.6) -- (-1.4,-1.4);

\draw      (0,0) -- (0.6,-0.6);
\draw [<-] (0.6,-0.6) -- (1.4,-1.4);

\fill (0,0) circle (0.8mm);

\draw (0,2.5) node {$J$};

\draw (-1.7,-1.7) node {$I$};

\draw (1.7,-1.7) node {$K$};

\draw (0.4, 0.3) node {$\eta$};

\end{tikzpicture}
\caption{Cubic vertex on the hyperbolic disk. Incoming  ($I$, $K$) and outcoming ($J$) momenta are constrained  by the  equilibrium condition. The radial vertex position is parameterized by $\eta = \cot^2 \rho$, where $\rho$ is the radial distance from the center.  } 
\label{vertex}
\end{figure}
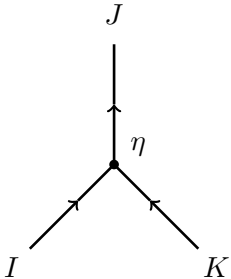

Any vertex of the geodesic network  on Fig.  \bref{bulk} connects three external and/or exchanged lines, where the vertex point is locally  given by three coinciding inner endpoints while outer endpoints are free, see Fig. \bref{vertex}. The vertex action for three distinct lines  has the form   
\be
\label{star}
S_{\star} =  \epsilon_I  \int^{\bullet \scriptscriptstyle }_{ \circ \scriptscriptstyle I} d \lambda\, L_I+\epsilon_J  \int^{\bullet \scriptscriptstyle }_{ \circ \scriptscriptstyle J} d \lambda\, L_J+ \epsilon_K  \int^{\bullet \scriptscriptstyle }_{ \circ \scriptscriptstyle K} d \lambda\, L_K\;, \qquad I\neq J\neq K\;,
\ee
where each term is the worldline action on the hyperbolic disk \eqref{disk} with the vertex point $\bullet \scriptscriptstyle$ and outer endpoints $\circ \scriptscriptstyle A$, where $A = I,J,K$. The least principle guarantees that the geodesic segments satisfy the  equilibrium condition at the vertex point  
\be
\label{conserv}
P^{(I)} + P^{(J)} + P^{(K)} = 0\;, 
\ee 
where $P_m^{(A)} = \partial L_{A} /\partial \dot{X}^m_{(A)}$  are canonical momenta of three particles with coordinates $X^m_{(A)}$, where $m = \rho, \phi$ and $A = I,J,K$. 

 The  equilibrium condition is conveniently parameterized by classical conformal dimensions $\epsilon_A$ and  angular parameters
\be
\label{sP}
s_A = \frac{|P^{(A)}_\phi|}{\alpha}\;.
\ee 
The parameter $s_A$ is an integration constant characterizing the form of a particular segment (see Appendix \bref{sec:B}). Since  $P^{(A)}_\phi  = \pm \alpha s_A$, where the overall sign depends on the direction of the flow,  we find that the  radial and angular  projections of \eqref{conserv} are given by 
\be
\label{eqcondi}
 \epsilon_I\sqrt{1-s_I^2 \eta} - \epsilon_J\sqrt{1-s_J^2  \eta}+\epsilon_K\sqrt{1-s_K^2 \eta}=0\;,
\ee
\be
\label{gens}
\epsilon_I s_I +\epsilon_J s_J - \epsilon_K s_K =0 \;,
\ee
where we expressed radial velocities through the vertex position according to  \eqref{velo}.

Using the linear relation \eqref{gens} we solve the radical equation \eqref{eqcondi}   as follows   
\be
\label{newetai}
\eta  =\frac{1 - \sigma_{\hspace{-1mm}{}_{IJ}}^2}{s_I^2 +s_J^2 - 2\sigma_{\hspace{-1mm}{}_{IJ}}  s_I s_J }\;, \qquad \text{where} \qquad \sigma_{\hspace{-1mm}{}_{IJ}} = \frac{\epsilon_I^2 +\epsilon_J^2 - \epsilon_K^2 }{2 \epsilon_I\epsilon_J}\;.
\ee
Remarkably, there are no other roots. Indeed, the  radical  equation \eqref{eqcondi} can be solved by isolating one of radicals on one side and  then squaring both sides. The resulting equation is linear in $\eta$. For example, the representation \eqref{newetai} is obtained by isolating the first radical in \eqref{eqcondi}. Other equivalent forms of the vertex position can be obtained by isolating the second and  third radicals,  
\be
\label{newetai2}
\eta  =\frac{1 - \sigma_{\hspace{-1mm}{}_{IK}}^2}{s_I^2 +s_K^2 - 2\sigma_{\hspace{-1mm}{}_{IK}}  s_I s_K }\;, \qquad \text{where} \qquad \sigma_{\hspace{-1mm}{}_{IK}} = \frac{\epsilon_I^2 +\epsilon_K^2 - \epsilon_J^2 }{2 \epsilon_I\epsilon_K}\;,
\ee
\be
\label{newetai3}
\eta  =\frac{1 - \sigma_{\hspace{-1mm}{}_{JK}}^2}{s_J^2 +s_K^2 + 2\sigma_{\hspace{-1mm}{}_{JK}}  s_J s_K }\;, \qquad \text{where} \qquad \sigma_{\hspace{-1mm}{}_{JK}} = \frac{\epsilon_J^2 +\epsilon_K^2 - \epsilon_I^2 }{2 \epsilon_J\epsilon_K}\;.
\ee
Note that equations \eqref{eqcondi} and \eqref{gens} are linear combinations of the same type terms, but with different signs. This is why $\sigma_{\hspace{-1mm}{}_{JK}}$ in \eqref{newetai3} has a different sign.   

The radicals in \eqref{eqcondi} impose restrictions on the radial vertex position
\be
0 \leq \eta \leq 1/s_A^2\;, \qquad  A = I,J,K\;.
\ee 
The region $\eta < 0$ is unphysical corresponding to imaginary values of the radial position. Examining   the region $\eta\geq 0$  we find out that the classical conformal dimensions necessarily satisfy the triangle inequalities.  

\begin{prop}
\label{prop1} The reality condition $\eta \geq 0$  is satisfied iff
\be
\label{triangle}
\ba{c}
\epsilon_I + \epsilon_J \geq \epsilon_K\;,
\\
\epsilon_I + \epsilon_K \geq \epsilon_J\;,
\\
\epsilon_J + \epsilon_K \geq \epsilon_I\;.

\ea
\ee  
\end{prop}

\noindent The proof is given in Appendix \bref{sec:prop1}.\footnote{The triangle inequalities  \eqref{triangle} are also satisfied by the vertex of two background heavy operators of dimension  $\epsilon_h \equiv  \epsilon_{n-1} = \epsilon_{n}$ and the perturbative heavy exchanged operator of dimension   $\tilde \epsilon_{n-3}$. In this case, $2\epsilon_h \gg \tilde \epsilon_{n-3}$ and $\tilde \epsilon_{n-3} \geq  0$. } From the triangle inequalities \eqref{triangle} it follows that the sigmas \eqref{newetai}--\eqref{newetai3} can be parameterized as cosines $\sigma_{AB} = \cos \gamma_{AB}$, where  $\gamma_{AB}$ is the angle between $\epsilon_A$ and $\epsilon_B$ sides of the triangle \eqref{triangle} in the space of conformal dimensions. Introducing  $s_{_{AB}}^2  = s_A^2 + s_B^2  \pm 2\sigma_{\hspace{-1mm}{}_{AB}}  s_A s_B $ according to \eqref{sss} we  represent the vertex position \eqref{newetai}--\eqref{newetai3} as 
\be
\label{ratios}
\ba{c}
\dps
\eta  = \left[\frac{\sin \gamma_{\hspace{-0mm}{}_{IJ}}}{s_{\hspace{-0mm}{}_{IJ}}}\right]^2 =\left[\frac{\sin \gamma_{\hspace{-0mm}{}_{IK}}}{s_{\hspace{-0mm}{}_{IK}}}\right]^2 =\left[\frac{\sin \gamma_{\hspace{-0mm}{}_{JK}}}{s_{\hspace{-1mm}{}_{JK}}}\right]^2\;.
\ea
\ee
In this form, it  is similar to the law of sines in planar trigonometry. Nevertheless, unlike the conformal dimensions, the  angular parameters $s_I, s_J, s_K$ do not form a triangle because,   generally, $s_{AB} \neq s_C$. It would be interesting to understand the role of the ratio $(\sin \gamma)/s$ as an invariant of the geometry of cubic vertices on the hyperbolic space. Here we just note that the triangle inequalities \eqref{triangle} are analogous to triangle inequalities satisfied by conformal dimensions of primary operators in the semiclassical limit of the DOZZ three-point correlation function \cite{Harlow:2011ny}. Note that  the triangle inequalities in the Liouville theory  are supplemented by the Seiberg bound \cite{Seiberg:1990eb} and the Gauss-Bonnet constraint \cite{Harlow:2011ny}. All together they guarantee that the Liouville field  solution is real.

\subsection{Dual geodesic network}
\label{sec:net}

Gluing together $n-3$ vertices $\eta_1, ...\,, \eta_{n-3}$ with endpoints attached to  $n-2$ boundary  points $w=(w_1, ... \,, w_{n-2})$ and the center of the disk  we obtain  the  network shown on Fig. \bref{bulk}. At this stage all segments are naturally divided into external and exchanged ones.  In what follows we use the condensed index $A = 1, ...\,, 2n-5$:
\be
\{A\} = \{i, \tilde j\}\;: \qquad  i = 1,...\,, n-2\;, \quad  \tilde j = 1, ...\,, n-3\;, 
\ee
to label $n-2$ external  and $n-3$ exchanged segments. The total worldline action is given by the sum of $n-3$ vertex actions \eqref{star}, {\it i.e.},  $S = \dps\sum_{m=1}^{n-3} S^{(m)}_\star$, where endpoints are connected to each other in such a way  to form  the network shown on Fig. \bref{bulk}. The resulting action reads
\be
\label{action}
S = \sum_{A=1}^{2n-5} \epsilon_A  \int^{\bullet \scriptscriptstyle  A}_{ \circ \scriptscriptstyle A} d \lambda\,  L_{A}\;,
\ee
where Lagrangian  $L_A$ describes $A$-th geodesic external/exchanged segment with endpoints $\circ \scriptscriptstyle  A$ and $\bullet \scriptscriptstyle A$ defined by the form of the network. Any vertex $\eta_{i}$,  $i=1,...\,, n-3$ joins lines with labels  $I = i+1, J =\tilde i, K = \widetilde{i-1}$, where, for convenience, we  equated  $\tilde 0 = 1$. 

Given that the  action functional $S$ is stationary we find the equilibrium conditions \eqref{conserv} at each vertex point  
\be
\label{equilib}
P_{(i+1)} + P_{(\tilde i)} + P_{(\widetilde{i-1})} = 0\;, \qquad i = 1,...\,, n-3\;,
\ee 
and  out-flowing momenta in all attachment points on the boundary and at the center of the disk,  
\be
\label{att}
P_{(A)} = \frac{\partial S}{\hspace{1mm}\partial X_{(A)}}\;, \qquad A = 1,...\,, n-3, \widetilde{n-3} \;, 
\ee
where the  last equality is assumed to be weak, {\it i.e.} the action $S$ is evaluated on-shell. 

All the attachment points have limiting radial positions, $\rho = \pi/2$ for the boundary points and $\rho=0$ for the center of the disk. It follows that radial components of \eqref{att} trivialize because the corresponding variation terms do not contribute to the variation  $\delta S$. The on-shell action of the network depends on angles of the boundary attachment points  and the center of the disk, $S =S(w, \phi_o)$. Using  \eqref{sP} we can represent non-vanishing components of \eqref{att} as angular gradients of the on-shell action
\be
\label{si}
s_i  =\frac{1}{\alpha \epsilon_i}\frac{\partial S}{\partial w_i}\;, \qquad i = 1,...\,, n-2\;,
\ee 
and 
\be
\label{s3}
\tilde s_{n-3}  =\frac{1}{\alpha \epsilon_i}\frac{\partial S}{\partial \phi_o }\;.
\ee
The angular coordinate $\phi_o$ of the center is arbitrary, then the derivative in \eqref{s3} vanishes, and, therefore,   
\be
\label{s30}
\tilde s_{n-3} = 0\;.
\ee 
It follows that the corresponding worldline  is radial. Recalling that the center contains the conical singularity we conclude that the radial fall of the outer exchanged particle is quite natural.
 
Angular components of the equilibrium conditions \eqref{equilib} can be  explicitly written as  
\be
\label{linrelss}
\ba{l}
\epsilon_{i} s_{i} +\tilde \epsilon_{i-1}\tilde s_{i-1} - \tilde \epsilon_{i-2}\tilde s_{i-2} = 0\;,
\qquad
i =2, ...\,, n-1\;.
\ea
\ee
We can then solve \eqref{linrelss} to obtain 
\be
\label{elimeta}
\tilde \epsilon_k  \tilde s_k = \epsilon_1 s_1  - \sum_{i=2}^{k+1} \epsilon_i s_i\;, \qquad  k =1, ...\,,n-3\;,
\ee
and therefore all exchanged momenta can be expressed in terms of the external ones. Taking $k = n-3$ we find out that the total outflowing angular parameter is zero. Indeed, in this case the right-hans side of \eqref{elimeta} is the sum of outflowing parameters at the boundary attachment points, while the left-hand side is the outflowing parameter  at the center of the disk \eqref{s30}. Thus, all external parameters  are linearly related as
\be
\label{slin}
-\epsilon_1 s_1  + \sum_{i=2}^{n-2} \epsilon_i s_i = 0\;.
\ee  

Choosing again  $I = \widetilde{k-2}, J = \widetilde{k-1}, K = k$ we find that radial components of the equilibrium conditions \eqref{equilib} can be  explicitly written as
\be
\label{eqrad}
\tilde \epsilon_{k-1} \sqrt{1 - \tilde s_{k-1}^2 \eta_{k-1}} - \tilde \epsilon_{k-2} \sqrt{1 - \tilde s_{k-2}^2 \eta_{k-1}}+ \epsilon_{k} \sqrt{1 - \tilde s_{k}^2 \eta_{k-1}} = 0\;.
\ee
Vertex positions $\eta_{k-1}$ are directly read off from the general formula \eqref{newetai}, namely 
\be
\label{vertvertI}
\eta_{k-1} = \frac{1-\sigma_k^2}{s_k^2 + \tilde s_{k-2}^2 - 2 \sigma_k s_k \tilde s_{k-2}}\;, \qquad \sigma_k = \frac{\epsilon_k^2 +\tilde\epsilon_{k-2}^2 - \tilde\epsilon_{k-1}^2 }{2 \epsilon_k\tilde\epsilon_{k-2}}\;, \qquad k = 2,...\,,n-2\;.
\ee
Equivalent representations of $\eta_{k-1}$ can be obtained using  \eqref{newetai2} and \eqref{newetai3}. 

\subsection{Angular balance condition}
\label{sec:balance}

\begin{figure}[H]
\centering
\begin{tikzpicture}[line width=1pt,scale=1]
\draw [line width=1pt,domain=-170:-20] plot ({3*cos(\x)}, {3*sin(\x)});

\foreach \a in {1,2,...,40}{
\draw (\a*360/40: 3.5cm) coordinate(N\a){};
\draw (\a*360/40: 3.7cm) coordinate(G\a){};
\draw (\a*360/40+5: 3.5cm) coordinate(D\a){};

\draw (\a*360/40: 4cm) coordinate(A\a){};
\draw (\a*360/40: 3cm) coordinate(K\a){};
\draw (\a*360/40: 2.5cm) coordinate(F\a){};
\draw (\a*360/40: 2cm) coordinate(L\a){};
\draw (\a*360/40: 1.5cm) coordinate(I\a){};
\draw (\a*360/40: 1.1cm) coordinate(J\a){};
\draw (\a*360/40: 1cm) coordinate(M\a){};
\draw (\a*360/40: 0.8cm) coordinate(C\a){};
;}


\draw[blue] plot [smooth, tension=1.0, line width=1pt] coordinates {(K27) (L28) (I30)};
\draw[blue] plot [smooth, tension=1.0, line width=1pt] coordinates {(K23) (L23) (M25)};


\draw  [blue, smooth, tension=1.0, line width=1pt, decorate, decoration = {snake, segment length = 2mm, amplitude=0.4mm}]  (I30) -- (J29) -- (M25);
\draw  [blue, smooth, tension=1.0, line width=1pt, decorate, decoration = {snake, segment length = 2mm, amplitude=0.4mm}]  (J19) -- (M25);

\draw  [blue, smooth, tension=1.0, line width=1pt, decorate, decoration = {snake, segment length = 2mm, amplitude=0.4mm}]  (I30) --  (L33);

\draw [red,dotted, line width=0.6pt] (0,0) -- (K27);
\draw [red,dotted, line width=0.6pt] (0,0) -- (K23);

\draw [red,dotted, line width=0.6pt] (0,0) -- (K30);
\draw [red,dotted, line width=0.6pt] (0,0) -- (K25);

\draw (N27) node {$w_{k-1}$};
\draw (N23) node {$w_{k}$};

\draw (N30) node {$\psi_{k-2}$};
\draw (N25) node {$\psi_{k-1}$};

\draw (L31) node {$\eta_{k-2}$};
\draw (I22) node {$\eta_{k-1}$};
\draw (C31) node {$\tilde s_{k-2}$};

\fill[blue] (I30) circle (0.8mm);
\fill[blue] (M25) circle (0.8mm);
\fill[blue] (K27) circle (0.8mm);
\fill[blue] (K23) circle (0.8mm);
\fill (0,0) circle (0.5mm);

\end{tikzpicture}

\caption{Angular separations. Dotted lines show angular positions $w_i$ and $\psi_i$   of the boundary attachment points and  vertices, respectively.} 
\label{angle}
\end{figure}
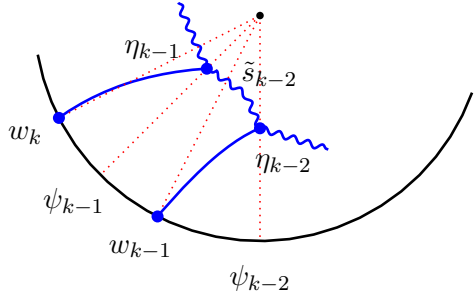

We study angular positions of the endpoints which define the limits of integration in the worldline action \eqref{action}. Let $\psi_{i}$  be  angles of  vertices $\eta_i$, $i = 1,...\,,n-3$ and $w_i$, $i = 1,...\,,n-2$ be angles of the boundary attachment points, see Fig. \bref{angle}. Generally, exchanged lines are stretched between  two neighboring vertices while external lines connect  vertices with boundary points. From Fig. \bref{angle} we find that the angular separation of the $i$-th external  segment is   
\be
\label{31avg}
\Delta \phi_i  = w_i - \psi_{i-1}\;, 
\qquad
i = 2, ...\,, n-2\;,
\ee
while the angular separation of the $i$-th exchanged  segment is    
\be
\label{32avg}
\Delta\tilde \phi_{i} = \psi_{{i+1}} - \psi_{i}\;,
\qquad
i = 1,...\,, n-4\;.
\ee
Both the rightmost and leftmost parts of the network on Fig. \bref{bulk} are different from the general pattern on Fig. \bref{angle}. Therefore, we identify $w_1 = \psi_0$ and hence  $\Delta \tilde \phi_0 = \psi_1 - \psi_0$ is the angular separation of the first external line. Also, $\Delta \tilde \phi_{n-3} = 0$ because the outer exchanged line is radial. From Fig. \bref{angle} we find that angular positions satisfy the balance equation
\be
\label{balance}
(w_k - w_{k-1}) + \Delta \phi_{k-1} = \Delta \phi_{k} +\Delta \tilde \phi_{k-2}\;,\qquad  k = 2,...\,, n-2\;.
\ee

\vspace{-30mm}

Angular separation of  the geodesic segment with two endpoints having radial and angular positions $(\phi^{'},\eta^{'})$ and $(\phi^{''},\eta^{''})$ and characterized by the angular parameter $s$ can be represented as \cite{Alkalaev:2015wia}
\be
\label{increment}
i \alpha(\phi^{''} - \phi^{'}) =\ln
\frac{\sqrt{1-s^2 \eta^{''}}-i s \sqrt{1+\eta^{''}}}
{\sqrt{1-s^2 \eta^{'}}-i s \sqrt{1+\eta^{'}}}\;,
\ee
cf. \eqref{angular}.
Boundary attachment points have $\eta = 0$, radial vertex positions $\eta_k$  are given by \eqref{vertvertI}. From the technical perspective, the logarithmic representation \eqref{increment} is well suitable for the bulk/boundary correspondence analysis of Section \bref{sec:duality} because  the conformal map from the complex plane to the boundary cylinder is also logarithmic \eqref{map}.

Substituting  \eqref{increment} into the angular balance equation \eqref{balance} we obtain the  system of radical equations  
\be
\label{newform}
e^{i\alpha (w_{k} - w_{k-1})} \frac{1-is_k}{1-is_{k-1}} \frac{D_{k-1}^-}{D_k^+} = 1\;, \qquad k = 2,...\,,n-2\;, 
\ee
where we introduced notation 
\be
\label{NEWangeqFant}
\dps
D_{k}^{-} = (\sqrt{1 - s_k^2\eta_{k-1}} - i s_k\sqrt{1+\eta_{k-1}})(\sqrt{1 - \tilde s_{k-1}^2\eta_{k-1}} - i \tilde s_{k-1}\sqrt{1+\eta_{k-1}})\;,\;\;\;\; k=1, ...\,, n-2\;,
\ee
\be
\label{NEWangeqFant2}
\dps
D_{k}^{+} = (\sqrt{1 - s_k^2\eta_{k-1}} - i s_k\sqrt{1+\eta_{k-1}})(\sqrt{1 - \tilde s_{k-2}^2\eta_{k-1}} - i \tilde s_{k-2}\sqrt{1+\eta_{k-1}})\;, \;\;\;\; k=2, ...\,, n-2\;.
\ee
The left-hand side of \eqref{newform} is unimodular so that the corresponding  real and imaginary parts are not independent. Thus, the  complex equations \eqref{newform} are equivalent to original real equations \eqref{balance}.\footnote{Exceedingly lengthy but equivalent representation of the angular equations was also given in  \cite{Alkalaev:2015wia}.}    

The angular balance equation \eqref{newform} is too complicated because of radicals depending on   vertex radial positions, which in turn  depend non-trivially on angular paprameters \eqref{vertvertI}. The roots of the system are yet unknown except for the simplest $n=4$ case \cite{Fitzpatrick:2014vua,Hijano:2015rla}, where the solution can be found exactly,  and the $n=5$ case \cite{Alkalaev:2015wia,Alkalaev:2015lca}, where only perturbative solutions are available.\footnote{Approximate  solutions in other conformal block channels were also  considered in \cite{Fitzpatrick:2014vua, Banerjee:2016qca}.}  In the next section we show that the dual network   equations \eqref{linrelss}, \eqref{vertvertI}, \eqref{newform} have  roots coinciding with those of the accessory parameter  equations \eqref{ccc}, \eqref{acc_sys}.

\section{Bulk/boundary correspondence}
\label{sec:duality}

The correspondence is most clear in the case of  2-point correlation function of heavy background operators. Two operators inserted in $z=1$ and $z = \infty$  produce an asymptotically $\ads$ geometry \eqref{metric} in the bulk with cylindrical conformal boundary. Angle deficit $\alpha$ in the metric \eqref{metric} is related to the background operator conformal dimension by \eqref{alpha}. This can be explicitly seen using the three-dimensional metric of asymptotically $\ads$ spacetime in the Banados form \cite{Banados:1998gg}, where the stress-energy tensor is taken to be the background value $T^{(0)}(z)$ \eqref{T0} (see \cite{Asplund:2014coa,Fitzpatrick:2015zha} for more details). Note that the CFT we started with is originally defined on the sphere. However, puncturing  the sphere twice we obtain the once punctured complex plane, which can be conformally mapped on the cylinder. Opposite ends of the resulting infinitely long cylinder correspond to the background operator insertions.\footnote{This topological consideration can be rigorously implemented within the Liouville theory with heavy insertions, see, \eg, \cite{ZZ}.} 

Let $(z, \bar z)$ be coordinates on the punctured complex plane and $(w, \bar w)$ be coordinates on the boundary cylinder. Then, the conformal map is given by 
\be
\label{map}
w  = i \ln (1-z)\;.
\ee
In our case, the boundary attachment points $w_i$  are located on a circle obtained by slicing the boundary cylinder, see Fig. \bref{bulk}. It follows that insertion points $z_i$ belong to the unit circle 
\be
\label{circle}
(z_i-1)(\bar z_i-1) = 1\;, \qquad  i = 1,...\,, n-2\;.
\ee
In particular, $z_1 = 0$ fixed by  projective $sl(2,\mathbb{C})$ isometry of the plane goes to $w_1=0$ fixed  by  $o(2)$ isometry transformation of the boundary circle.

We argue that  the correspondence can be shown without knowing  explicit expressions for conformal blocks/geodesic lengths. We have seen that both  classical the conformal block and the geodesic length are defined through auxiliary parameters subjected to algebraic equation systems. It is natural to question whether two equation systems are equivalent provided that the conformal block and geodesic length are related by the conformal map. We might hope that they are literally the same. However, it turns out that the equivalence is weaker. It is claimed only that being generally different the two systems have the same roots. This section addresses such an equivalence in the $n$-point case. 

The correspondence between a CFT with two background operators in the infinite central charge limit and dual geodesic networks on the angle deficit background  claims that the perturbative classical $n$-point block \eqref{ccb5}  and the on-shell worldline action of the dual network \eqref{action} be related to each other as
\be
\label{fS}
f(z) = S(w)+ i\sum_{k=1}^{n-2}\epsilon_k w_k\;. 
\ee
Here, the last term results from conformal transformations \eqref{map} of $n-2$  perturbative heavy operators. Indeed, the corresponding correlation function gains the factors $(dw/dz)^{-\epsilon_i}$ in the insertion points, while the quantum conformal block is exponentiated to give the classical conformal block \eqref{ccb5}.

\begin{prop} 
\label{propro}
Given \eqref{map} and \eqref{fS} the accessory and angular parameters are related as 
\be
\label{cs}
c_k = \epsilon_k \, \frac{1\pm i \alpha s_k}{1-z_k}\;,\qquad k = 1,...\,,n-2\;,
\ee
with the convention that "$-$" at  $k=1$ and  "$+$" at  $k \neq 1$. 
\end{prop}
To prove the proposition  we recall that both  accessory parameters of  perturbative heavy operators and  angular parameters of the corresponding external worldlines are uniformly  defined  as the gradients,
\be
\label{again}
c_k  = \frac{\partial f}{\partial z_k}\;, \qquad \qquad s_k  =\frac{1}{\alpha \epsilon_k}\frac{\partial S}{\partial w_k}\;, \qquad\qquad k = 1, ...\,, n-2\;,
\ee
cf. \eqref{acs} and \eqref{si}. 
Then, the relation \eqref{cs} directly follows  from  the conformal transformation \eqref{map} applied to the systems \eqref{again} supplemented with  the change  \eqref{fS}. 

To complete the proof of the correspondence we show that the algebraic equation systems imposed on the accessory and angular parameters have the same roots. To this end, we find the same  structures on both sides and partially solve the bulk system to show that the resulting equations coincide with the boundary equations.

\subsection{Weak equivalence}
\label{sec:potential}

On the formal level, the problem is as follows. We consider a potential vector field 
\be
\label{pot}
A_i(x) = \frac{\partial U(x)}{\partial x^i}\;, \qquad i = 1,...\,, n-2\;,
\ee
on a manifold with $n-2$ local coordinates  $x^i$ and  potential $U(x)$. Let the vector components $A_i$ be subjected to algebraic equations given implicitly by 
\be
\label{lag}
C^{(N)}_{\alpha}(A,B) = 0\;, \qquad \alpha =1, ...\,, N\;,
\ee 
where $B_k$ are possible auxiliary variables, $k = 1,...\,, N-(n-2)$. The coefficients in \eqref{lag} may explicitly depend on coordinates $x$ and some additional parameters. We assume that the algebraic system \eqref{lag} is non-degenerate and therefore the auxiliary variables can be completely expressed in terms of the potential vector field components, $B = B(A)$.       

We consider two potential vector field systems defined by two different sets  
\begin{align}
&\{x, U(x), A(x), B(x), C^{(N)}\}\;, \label{W1}\\
&\{y, \widetilde U(y), \widetilde A(y), \widetilde B(y), \widetilde C^{(\widetilde N)}\}\label{W2}\;.
\end{align} 

\begin{definition} Two systems  \eqref{W1}, \eqref{W2} are called weakly equivalent if 
the  implicit relations  ($\alpha = 1, ...\,, N$ and $\tilde \beta  = 1, ... , \widetilde N$) 
\be
C_{\alpha}(A,B) = 0\;, \qquad C_{\tilde \beta}(\widetilde A,\widetilde B) = 0\;,
\ee 
have at least one common root $\{A_i^0(x)\} \to  \{\widetilde A^0_i(y)\}$ under transformations 
\be
x \to y\;,\qquad U(x) \to \widetilde U(y)\;.
\ee  
\end{definition}

In our case, the boundary system has no B-type variables which are characteristic of the bulk system. This is quite natural from the AdS/CFT perspective in the sense that not all  bulk degrees of freedom are fundamental. Integrating out the local degrees of freedom identified here with B-type variables we are left with A-type variables which are fundamental boundary variables.      

\subsection{Accessory equations }
\label{sec:accsys}

Fixing $\epsilon_{n-1} = \epsilon_n$ we summarize the accessory  equations \eqref{ccc} and \eqref{acc_sys}, 
\begin{align}
&c_1 = -\sum^{n-2}_{i=2}\big[c_i(1-z_i)- \epsilon_i\big] +\epsilon_1 \;, 
\label{mon1}\\
&(I_{++}^{(k)})^2 + I_{-+}^{(k)}I_{+-}^{(k)} =  4\pi^2 \, \tilde\epsilon^2_{k}\;,
\qquad k = 1,...\,, n-3\;, \label{mon2}
\end{align}
where independent variables are  $c_1, ...\,, c_{n-2}$, while $I_{\pm\pm}^{(k)}$ are given by  \eqref{I+-}, \eqref{I++}. In total, there are $(n-2)$ equations for $(n-2)$ variables. 

Given the conformal map \eqref{map} we introduce the following notation 
\be
\label{AA}
A_p = a_p(1+is_p)\;,\qquad \bar A_p= \bar{a}_p(1-is_p)\;,\qquad \text{where}\qquad a_p = (1-z_p)^\alpha\;.
\ee
Insertions points satisfy the unit circle constraint \eqref{circle} so that $\bar{a}_p = 1/a_p$. Moreover, in order to simplify the accessory equations we recover the exchanged angular momenta   expressed via the external ones \eqref{elimeta}. As a result, we rewrite the accessory equations in terms of the external and exchanged momenta.  

\begin{prop}
\label{monprop}
Accessory equations  \eqref{mon1} and \eqref{mon2} are equivalently rewritten in terms of the angular parameters as 
\begin{align}
&\epsilon_1 s_1  - \sum_{i=2}^{n-2} \epsilon_i s_i = 0\;, \label{friday} \\
&\re \left[2\pi i \bar A_{k+1} I_{+-}^{(k-1)}\right] = 4 \pi^2 \tilde\epsilon_{k-1}\left(s_{k+1}\tilde s_{k-1}+ \sigma_{k+1}\right)\label{monprop1}
\;,
\end{align}
where $ k = 1, ...\,,n-3$, and 
\be
\label{monprop2}
I_{+-}^{(k-1)} = 2 \pi i \Big(\epsilon_1 \bar A_1 + \sum_{p=2}^{k} \epsilon_p A_p \Big) \;,
\ee
and $\sigma_k$ is given by \eqref{vertvertI},  $\tilde s_k$ is given by \eqref{elimeta}.
\end{prop} 

\noindent The proof is given in Appendix \bref{sec:monprop}.

\subsection{Momentum  equations}
\label{sec:ang}

Using notation \eqref{AA} we summarize the momentum equations consisting of the equilibrium conditions  \eqref{linrelss}, \eqref{eqrad}, and the angular balance condition \eqref{newform} as 
\begin{align}
&\epsilon_{i} s_{i} +\tilde \epsilon_{i-1}\tilde s_{i-1} - \tilde \epsilon_{i-2}\tilde s_{i-2} = 0\;,
\qquad
i =2, ...\,, n-1\;, \label{mond1}\\
& \tilde \epsilon_{k-1} \sqrt{1 - \tilde s_{k-1}^2 \eta_{k-1}} - \tilde \epsilon_{k-2} \sqrt{1 - \tilde s_{k-2}^2 \eta_{k-1}}- \epsilon_{k} \sqrt{1 - \tilde s_{k}^2 \eta_{k-1}} = 0\;,\;\;\;\; k = 2,...\,, n-2\;, \label{mond2}\\
&  D_k^+ = \frac{\bar A_{k+1}}{\bar A_{k}} D^-_{k-1}\;, \qquad k = 2,...\,,n-2\;, \label{mond3}
\end{align}
where independent variables are angular parameters $s_1, ...\,, s_{n-2}$, $\tilde s_1, ...\,, \tilde s_{n-3}$, and vertex positions $\eta_1, ...\,, \eta_{n-3}$, while $D^{\pm}_k$ are given by \eqref{NEWangeqFant}, \eqref{NEWangeqFant2}. In total, there are $(3n-8)$ equations for $(3n - 8)$ variables.  

In the bulk there are two types of redundant variables compared to those on the boundary:  exchanged angular parameters $\tilde s_k$ and vertex radial positions $\eta_k$. These are $B$-type variables of Section \bref{sec:potential}. The A-type variables  in this case are accessory parameters and external momenta. It follows that a straightforward way to compare two descriptions is to express $\eta_k$ in terms of $s_i$ and $\tilde s_j = \tilde s_j(s_i)$ by \eqref{mond1}, \eqref{mond2} using  \eqref{vertvertI} and then substitute them  into the angular balance equations \eqref{mond3}. However, the resulting equations depending only on $s_i$ turn out to be very complicated. In what follows we partially solve the momentum equations that drastically simplifies the analysis of the correspondence.    

To this end we notice that functions $D_k^{\pm}$ are building blocks of the angular balance equation \eqref{mond3}. We study their properties given that the equilibrium equations \eqref{mond1} and \eqref{mond2} are satisfied and find out that they are proportional to the contour integrals $I_{+-}^{(k)}$. Two lemmas are in order. 

\begin{lemma} 
\label{lemma}
Given  the equilibrium conditions \eqref{mond1} and \eqref{mond2}, the functions $D_k^-$ \eqref{NEWangeqFant} and $D_k^+$ \eqref{NEWangeqFant2} are linearly dependent  
\be
\label{DD}
\tilde \epsilon _{k-1} D^-_k  - \tilde \epsilon_{k-2} D^+_k = \epsilon_k A_k \bar A_k\;, 
\qquad k = 2,...\,, n-2\;.
\ee

\end{lemma}
\noindent The proof is given in Appendix \bref{sec:lemma}.

\begin{lemma}
\label{lemmaRE} Given the equilibrium conditions \eqref{mond1} and \eqref{mond2}, the real part of $D^+_k$ can be chosen in the form 
\be
\label{ReD+}
\re D_k^+  = -( s_k\tilde s_{k-2}  + \sigma_k)\;, \qquad k=2,...\,, n-2\;, 
\ee
where $\sigma_k$ is given by \eqref{vertvertI}.
\end{lemma}

\noindent The proof is given in Appendix \bref{sec:lemmaRE}.

\vspace{2mm}

The following proposition states that the momentum equations can be reformulated in terms of  new variables $D^{\pm}_k$.  
\begin{prop} 
\label{propNEW}
Equations \eqref{mond1} -- \eqref{mond3} can be rewritten as 
\begin{align}
& \epsilon_{i} s_{i} +\tilde \epsilon_{i-1}\tilde s_{i-1} - \tilde \epsilon_{i-2}\tilde s_{i-2} = 0\;,\qquad i =2, ...\,, n-1\;,\label{wed1}\\
& \re D_{k+1}^+  = -( s_{k+1}\tilde s_{k-1}  + \sigma_{k+1})\;, \qquad k=1,...\,, n-2\;,\label{wed2}
\end{align}
where $D^-_k$ and $D^{+}_{k+1}$, $k=1,...\,, n-2$ satisfy conditions
\begin{align}
& \tilde \epsilon _{k-1} D^-_k  - \tilde \epsilon_{k-2} D^+_k = \epsilon_k A_k \bar A_k\;,\qquad k = 1,... \;,\label{relD1}\\
&D_k^+ = \frac{\bar A_{k+1}}{\bar A_{k}} D^-_{k-1}\;, \qquad k = 2,...\,,n-2\;. \label{relD2}
\end{align}
\end{prop}
The proof directly follows  from Lemmas \bref{lemma} and \bref{lemmaRE}. Here, the vertex radial positions have been  eliminated by means of introducing new independent variables  $D^{\pm}_k$ satisfying  relations  \eqref{relD1} and \eqref{relD2}. In this form the momentum equations are similar to the accessory  equations of Proposition \bref{monprop}.

\subsection{Comparing two systems}
\label{sec:vs}

Putting Propositions \bref{monprop} and \bref{propNEW} together we find out that the two equation systems can be weakly equivalent if the contour integrals $I_{-+}^{(k-1)}$ and   functions $D_k^+$ are related as $\bar A_{k+1} I_{-+}^{(k-1)}  \sim D_k^+$. On the other hand, we might conclude that this cannot be the case because given the linear relation \eqref{cs} the contour integrals are quadratic functions of accessory parameters, while $D^{\pm}_k$  are composite functions with radicals and rational functions of momenta, cf. \eqref{NEWangeqFant}, \eqref{NEWangeqFant2}. However, from Proposition \bref{propNEW} it follows that $D^{\pm}_k$ are now independent variables subjected to linear equations \eqref{relD1} and \eqref{relD2} with coefficients at most quadratic in $A_k$ and $\bar A_k$. Below we explicitly solve these linear relations. 

First, in Section \bref{sec:net} we showed that the relation \eqref{wed1} can be solved in terms of external parameters to give \eqref{friday}, see \eqref{linrelss}--\eqref{slin}.  

Second, recalling that the lowest indices can be identified  as $\tilde 0 = 1$ implying  $\tilde \epsilon_{0} \equiv \epsilon_1 $ we find that equations \eqref{relD1} are explicitly written as
\be
\label{fin1}
\epsilon_1 D_1^- = \epsilon_1 A_1\bar A_1\;, 
\qquad 
\tilde \epsilon_1 D_2^-  = \epsilon_1 D_2^+ + \epsilon_2 A_2 \bar A_2\;,
\qquad
\tilde \epsilon_2 D_3^-  = \epsilon_2 D_3^+ + \epsilon_3 A_3 \bar A_3\;,  
\qquad \cdots\quad \;,
\ee
while equations \eqref{relD2} are 
\be
\label{fin2}
D_2^+ = \frac{\bar A_2}{\bar A_1} D_1^-\;, 
\qquad 
D_3^+ = \frac{\bar A_3}{\bar A_2}D_2^-\;, 
\qquad
D_4^+ = \frac{\bar A_4}{\bar A_3}D_3^-\;, 
\qquad \cdots\quad  \;.
\ee
Solving    \eqref{fin1} and \eqref{fin2} recursively we write down the general solution 
\begin{align}
D^+_{k+1} = \frac{1}{\tilde \epsilon_{k-1}}\bar A_{k+1} (\epsilon_1 \bar A_1 + \sum_{p=2}^{k} \epsilon_p A_p)\;, \qquad D^-_{k}  = \frac{1}{\tilde \epsilon_{k-1}}\bar A_{k} (\epsilon_1 \bar A_1 + \sum_{p=2}^{k} \epsilon_p A_p)\;. 
\end{align}
Now, recalling \eqref{monprop2} we  find the relation 
\be
D^+_{k+1} = \frac{1}{2\pi i \tilde \epsilon_{k-1}}\bar A_{k+1}\, I_{+-}^{(k-1)}\;,
\ee
which finally proves 
\begin{prop}
\label{equiv}
The accessory equations \eqref{mon1} and \eqref{mon2} are weakly  equivalent to the momentum equations \eqref{wed1} and \eqref{wed2}. 

\end{prop}

The weak equivalence assumes that the bulk system has  more roots compared to the boundary system. Our choice of roots is hidden in Lemma \bref{lemmaRE}. In Appendix \bref{sec:lemmaRE} we show that there is the other possible value of $\re D_k^+$. However, the corresponding root \eqref{2root} does not give rise to known dual counterparts. In other words, it seems that there are admissible bulk configurations which cannot be realized through the classical conformal block.  

\section{Conclusion}
\label{sec:con}

In this work we showed that $n$-point classical conformal blocks in the heavy-light approximation are equal (modulo the conformal map) to the lengths of dual geodesic networks for any $n$. To this end we  reformulated both bulk/boundary systems as the potential vector field equations, where vector components are subjected to the algebraic equations. Given the conformal map from the complex plane to cylinder we  demonstrated  that both algebraic systems share the same roots. This guarantees the correspondence even though explicit block and dual length functions are not presently known (except for various approximations). Moreover, using the notion of the weak equivalence we  showed that the roots of the boundary system is the subset in the roots of the bulk system. The role of  extra roots in the bulk  will be studied elsewhere.  

A possible future direction is to apply our technique to a semiclassical CFT on higher genius Riemann surfaces starting with the torus. The classical toroidal blocks were considered, \eg, in  \cite{Menotti:2010en,Piatek:2013ifa}, while their holographic interpretation in the heavy-light semiclassical  approximation was proposed in \cite{Alkalaev:2016ptm}. However, the perturbative  monodromy approach for toroidal and  higher genius CFTs and its holographic interpretation similar to that  in the spherical case have not yet been elaborated.   

Also, the semiclassical correspondence considered in this paper can be extended by including $1/c$ corrections. The 4-point case was studied in \cite{Beccaria:2015shq,Fitzpatrick:2015dlt,Fitzpatrick:2016ive,Chen:2016cms,Fitzpatrick:2016mjq}. 
It would be interesting to understand how our results for $n$-point  blocks connect with going beyond the leading $1/c$ order.

\vspace{7mm} 

\noindent \textbf{Acknowledgements.} I am grateful to Vladimir Belavin  for useful exchanges. 
I thank Bengt Nilsson and Aritra Banerjee for a stimulating discussion.    
Also, I would like to thank the Munich Institute for Astro- and Particle Physics (MIAPP) for hospitality during  the programme "Higher-Spin Theory and Duality" (May 2016).
The work  was supported by RFBR grant No 14-02-01171.


\appendix

\section{Technical details}
\label{sec:A}

\subsection{Worldlines in the angle deficit  geometry}
\label{sec:B}

Here we collect main formulas of the worldline formulation on the angle deficit space  \cite{Hijano:2015rla,Alkalaev:2015wia}. The worldline action associated to the metric \eqref{metric} is given by 
\be
\label{OPA}
S = \epsilon\int_{\lambda^{'}}^{\lambda^{''}} d\lambda \, \sqrt{\alpha^2 \sec^2 \rho\, \dot{t}^2+\alpha^2 \tan^2\rho\, \dot{\phi}^2+\sec^2 \rho\, \dot{\rho}^2}\;.
\ee 
It is reparameterization invariant and therefore the evolution parameter can be conveniently chosen such that the proper velocity is unit. It follows that the Lagrangian function \eqref{OPA} is unit $\alpha^2\sec^2 \rho\, \dot{t}^2+\alpha^2 \tan^2\rho\, \dot{\phi}^2+\sec^2 \rho\, \dot{\rho}^2 = 1$ 
and the action is simply the geodesic length of a segment stretched between endpoints $\lambda^{'}$ and $\lambda^{''}$.  Choosing the constant time slice $t=0$ we find that the normalization condition is cast into the form 
\be
\label{geod}
\frac{p_{\phi}^2}{\alpha^2}\cot^2\rho + \sec^2\rho \,\dot{\rho}^2= 1\;. 
\ee 
where $p_\phi$ is the conserved angular momentum associated with the cyclic coordinate $\phi$. 
A tricky point here is that we can avoid solving the geodesic equations of motion explicitly because the normalization condition \eqref{geod} is sufficient to express a proper parameter $\lambda$ as a function of radius and  angular momentum. From \eqref{geod} we find the radial velocity 
\be
\label{velo}
\dot \rho = \pm \cos\rho\, \sqrt{1 - \frac{p_\phi^2}{\alpha^2}\cot^2 \rho}\;\;.
\ee
Recalling $\dot \rho = d\rho/d \lambda $ we find that equation \eqref{velo} can be directly integrated to obtain the on-shell value of $S$ on the hyperbolic disk
\be
\label{lambda}
S = \ln \frac{ \sqrt{\eta}}{\sqrt{1+\eta} +  \sqrt{1 - s^2 \eta}}\,\Bigg|_{\eta^{'}}^{\eta^{''}}\;,
\ee
where $\eta^{'} = \cot^2 \rho^{'}$ and $\eta^{''} = \cot^2 \rho^{''}$ are initial/final radial positions. Here we used parameter $s = |p_\phi|/\alpha$ which is an integration constant describing  the shape of a geodesic segment, cf. \eqref{sP}.

Using $p_\phi = g_{\phi\phi} \dot \phi$ and recalling that the angular momenta are integration  constants we find that the angle increment of the geodesic segment is given by 
\be
\label{angular}
\Delta \phi = \pm \frac{p_\phi}{\alpha^2}\dps\int_{\rho^{'}}^{\rho^{''}} \frac{d\rho \cos \rho}{\sin^2 \rho (1 - \frac{p_\phi^2}{\alpha^2}\cot^2 \rho)^{1/2}}\;.
\ee 
Taking the integral we arrive at the logarithm representation  \eqref{increment}.

\subsection{Proof of Proposition \bref{prop1}  }
\label{sec:prop1}

To prove the proposition we introduce the classical fusion polynomial  \cite{Alkalaev:2015wia}
\be
\Pi_{IJK} = (\epsilon_I + \epsilon_J+ \epsilon_K)(-\epsilon_I + \epsilon_J+ \epsilon_K)(\epsilon_I - \epsilon_J+ \epsilon_K)(\epsilon_I + \epsilon_J- \epsilon_K)\;.
\ee
The  polynomial has two main properties \footnote{Note that the classical fusion polynomial is just the Heron's function defining the area of the triangle
$\text{Area}_{\Delta}(\epsilon_I, \epsilon_J, \epsilon_K) = \frac{1}{4}\sqrt{\Pi(\epsilon_I, \epsilon_J, \epsilon_K)}$ in the space of conformal dimensions.}
\be
\label{P1}
\Pi_{IJK} = \Pi_{IKJ}=\Pi_{JKI}\;,
\ee
and 
\be
\label{P2}
\Pi_{IJK} = 4 \epsilon_I^2 \epsilon_J^2 (1 - \sigma^2_{\hspace{-1mm}{}_{IJ}})\;,
\qquad 
\Pi_{IKJ} = 4 \epsilon_I^2 \epsilon_K^2 (1 - \sigma^2_{\hspace{-1mm}{}_{IK}})\;,
\qquad 
\Pi_{JKI} = 4 \epsilon_J^2 \epsilon_K^2 (1 - \sigma^2_{\hspace{-1mm}{}_{JK}})\;,
\ee
where the sigmas are given by \eqref{newetai}--\eqref{newetai3}. We introduce 
\begin{align}
&\;s_{IJ} = s_I^2 +s_J^2 - 2\sigma_{\hspace{-1mm}{}_{IJ}}  s_I s_J\;,\nonumber\\
&s_{IK} = s_I^2 +s_K^2 - 2\sigma_{\hspace{-1mm}{}_{IK}}  s_I s_K\;,\label{sss}\\
&s_{JK} = s_J^2 +s_K^2 + 2\sigma_{\hspace{-1mm}{}_{JK}}  s_J s_K\;.\nonumber
\end{align}
The vertex positions \eqref{newetai}, \eqref{newetai2}, \eqref{newetai3} are then given by
\be
\label{etas}
\eta = \left(\frac{1}{2\epsilon_I \epsilon_J}\right)^2\,\frac{\Pi_{IJK}}{s_{IJ}}\;,
\qquad
\eta = \left(\frac{1}{2\epsilon_I \epsilon_K}\right)^2\,\frac{\Pi_{IKJ}}{s_{IK}}\;,
\qquad
\eta = \left(\frac{1}{2\epsilon_J \epsilon_K}\right)^2\,\frac{\Pi_{JKI}}{s_{JK}}\;.
\ee 

Assume that $\epsilon_I+ \epsilon_J < \epsilon_K$. Squaring this inequality and using \eqref{P2} we find that $\sigma_{\hspace{-1mm}{}_{IJ}}< -1$, and $\Pi_{IJK}<0$. On the other hand, the angular momenta are non-negative $s_{I,J,K}\geq 0$,  so that if $\sigma_{\hspace{-1mm}{}_{IJ}}< -1$ then $s_{IJ}\geq 0$. Using for $\eta$ the first representation in  \eqref{etas} 
we conclude that $\eta <0$. It follows that $\epsilon_I+ \epsilon_J \geq \epsilon_K$ is  necessary to have real vertex positions.   

The second triangle inequality is proved along the same lines. Suppose that $\epsilon_I+ \epsilon_K < \epsilon_J$. Using for $\eta$ the second representation in  \eqref{etas}
we conclude that $\eta \geq 0$ provided  $\epsilon_I+ \epsilon_K \geq \epsilon_J$. 

Now, suppose  that $\epsilon_J+ \epsilon_K < \epsilon_I$. In this case, it is convenient to represent $\eta$ by  the third formula in \eqref{etas}, 
where $\sigma_{\hspace{-1mm}{}_{JK}}$ in $s_{JK}$ has an opposite sign compared to $s_{IJ}$ and $s_{IK}$, cf.  \eqref{sss}. Substituting the other two established triangle inequalities $\epsilon_I+ \epsilon_J \geq \epsilon_K$ and $\epsilon_I+ \epsilon_K \geq \epsilon_J$ into $\Pi_{JKI}$ we find that $\Pi_{JKI}\geq 0$. On the other hand, squaring $\epsilon_J+ \epsilon_K \geq \epsilon_I$ yields $\sigma_{\hspace{-1mm}{}_{JK}}> 1$, hence $s_{JK} \geq 0$. 

We conclude that having the triangle inequalities \eqref{triangle} is a necessary condition for $\eta\geq 0$.
 
\subsection{Proof of Proposition \bref{monprop}}
\label{sec:monprop}

Recalling that the heavy background dimensions are equal, $\epsilon_{n-1} = \epsilon_n$, we find that \eqref{ccc} takes the form  
\be
\label{modrel}
c_1 - \epsilon_1 = -\sum^{n-2}_{i=2}\big[c_i(1-z_i)- \epsilon_i\big]\;.
\ee
Under the correspondence map \eqref{cs} it transforms to the relation between the external angular parameters \eqref{slin}, 
\be
\epsilon_1 s_1  - \sum_{i=2}^{n-2} \epsilon_i s_i = 0\;.
\ee   

Now, using \eqref{cs} the contour integrals \eqref{I+-}--\eqref{I++} can be expressed in terms of external angular parameters $s_k$, $k=1,...\,,n-3$. Using  notation \eqref{AA} we find that the contour integrals are given by 
\begin{align}
\label{coefmom}
&I_{+-}^{(k)} = 2\pi i \left[\epsilon_1 \bar A_1 + \sum_{p=2}^{k+1} \epsilon_p A_p \right] \equiv  I_{+-}^{(k-1)}+ 2\pi i \epsilon_{k+1} A_{k+1} \;,
\\
&I_{-+}^{(k)} = 2\pi i \left[\epsilon_1 A_1 + \sum_{p=2}^{k+1} \epsilon_p \bar A_p \right] \equiv I_{-+}^{(k-1)}+ 2\pi i \epsilon_{k+1} \bar A_{k+1} \;,
\\
&I_{++}^{(k)} = \;2\pi  \left[ - \epsilon_1 s_1 + \sum_{p=2}^{k+1} \epsilon_p s_p \right] \equiv I_{++}^{(k-1)}+ 2\pi i \epsilon_{k+1} s_{k+1}\label{coefmom2}\;,
\end{align}
with the convention that $I_{+-}^{(0)} \equiv 2\pi i \epsilon_1 \bar A_1$ and $I_{++}^{(0)} \equiv   - 2\pi  \epsilon_1 s_1$.
In particular, as a consistency check  we find  from \eqref{slin} that  $I_{++}^{(n-3)} = 0$, see our comment below \eqref{I++}.

The accessory equations \eqref{mon1} are then represented as 
\be
\label{eqq}
\Big(-\epsilon_1 s_1 + \sum_{p=2}^{k+1} \epsilon_p s_p\Big)^2 - \Big(\epsilon_1 \bar A_1 + \sum_{p=2}^{k+1} \epsilon_p A_p \Big)\Big(\epsilon_1 A_1 + \sum_{p=2}^{k+1} \epsilon_p \bar A_p \Big) + \tilde \epsilon_k^2 = 0\;,
\ee
where $k = 1,2,...\,, n-3$. Equations \eqref{eqq} at $k$ and $k-1$ are related as follows. We substitute the right-hand side equalities of \eqref{coefmom}--\eqref{coefmom2} into \eqref{eqq} and find that 
\be
4 \pi s_{k+1} I_{++}^{(k-1)} + 2 \pi i \Big(A_{k+1}I_{-+}^{(k-1)} + \bar A_{k+1} I_{+-}^{(k-1)}\Big) = 8 \pi^2 \tilde \epsilon_{k-1} \sigma_{k+1}\;,
\ee 
where the sigma is given by \eqref{vertvertI}. Furthermore, we note that $I_{++}^{(k-1)}$ can be expressed in terms of the exchanged angular parameters by means of the relation \eqref{elimeta}, namely, $I_{++}^{(k-1)} =  - 2\pi \tilde \epsilon_{k-1} \tilde s_{k-1}$. Finally, we get 
\be
2\pi i \Big(A_{k+1} I_{-+}^{(k-1)}+ \bar A_{k+1} I_{+-}^{(k-1)}\Big) = 8 \pi^2 \tilde \epsilon_{k-1}(\tilde s_{k-1} s_{k+1}+ \sigma_{k+1})\;.
\ee 
Noting that $(I_{-+}^{(k)})^* =  - I_{+-}^{(k)}$, where $*$ denotes the complex conjugation and using  $\re\, x = (x+x^*)/2$  we conclude that the resulting equation is  exactly \eqref{monprop1}, while \eqref{monprop2} is \eqref{coefmom}.  

\subsection{Proof of Lemma \bref{lemma}}
\label{sec:lemma}

Following  definition \eqref{NEWangeqFant} and \eqref{NEWangeqFant2} we consider the difference
\be
\label{differ}
\tilde \epsilon_{k-1} D_k^-  - \tilde \epsilon_{k-2} D_k^+ \equiv  \left(\sqrt{1-s_k^2 \eta_{k-1}} - i s_k \sqrt{1+\eta_{k-1}}\right) \Delta B_{k}\;,
\ee
where 
\begin{align}
&\Delta B_k = \tilde \epsilon_{k-1} (\sqrt{1-\tilde s_{k-1}^2 \eta_{k-1}} - i \tilde s_{k-1} \sqrt{1+\eta_{k-1}})\nonumber\\
&\hspace{8mm}- \tilde \epsilon_{k-2} (\sqrt{1-\tilde s_{k-2}^2 \eta_{k-1}} - i \tilde s_{k-2} \sqrt{1+\eta_{k-1}})=\\
& \hspace{8mm} = (\tilde \epsilon_{k-1} \sqrt{1-\tilde s_{k-1}^2 \eta_{k-1}}  - \tilde \epsilon_{k-2} (\sqrt{1-\tilde s_{k-2}^2 \eta_{k-1}} ) + i (\tilde \epsilon_{k-2}\tilde s_{k-2}-\tilde \epsilon_{k-1}\tilde s_{k-1})\;.\nonumber
\end{align}
The real and imaginary parts here are given by the equilibrium conditions \eqref{mond2} and \eqref{mond1}, respectively. It follows that  
\be
\Delta B_k = \epsilon_k \left(\sqrt{1+s_k^2 \eta_{k-1}}+ i \sqrt{1+\eta_{k-1}}\right)\;,
\ee
which is modulo $\epsilon_k$ is the complex conjugate of the first factor in \eqref{differ}. Using \eqref{circle} and \eqref{AA} we find that the absolute value of this number is $1+s_k^2 = A_k \bar A_k$. Thus, \eqref{DD} holds true.

\subsection{Proof of Lemma \bref{lemmaRE}}
\label{sec:lemmaRE}

Here we partially solve the momentum equations. To this end, we solve \eqref{NEWangeqFant2} in terms of the radial vertex  positions 
\be
\label{angvert2}
\eta_{k-1} = \frac{1 - \left(\re D_k^++s_k\tilde s_{k-2}\right)^2}{s_k^2 + \tilde s_{k-2}^2 + 2 s_k \tilde s_{k-2}\left(\text{Re} D_k^++s_k\tilde s_{k-2}\right)}\;,
\ee 
where $D_k^+$ is not arbitrary but restricted by the balance equation \eqref{newform}. 
The real part is conveniently defined as
$\text{Re} D^+_k =(D^+_k+D^{+*}_k)/2$, where $*$ denotes the complex conjugation.   
There are no other  solutions since equations \eqref{NEWangeqFant2} are solved by squaring the radicals: squaring twice we obtain  linear in $\eta_k$ equations (see our comments below \eqref{newetai}). 

Equating two different but equivalent representations \eqref{vertvertI} and \eqref{angvert2} of the vertex radial position $\eta_{k-1}$ we find that  the real part of $D_k^{+}$ satisfies the quadratic equation 
\be
\label{Red}
\frac{1-\sigma_k^2}{s_k^2 + \tilde s_{k-2}^2 - 2 \sigma_k s_k \tilde s_{k-2}} = \frac{1 - \left(\text{Re} D_k^++s_k\tilde s_{k-2}\right)^2}{s_k^2 + \tilde s_{k-2}^2 + 2 s_k \tilde s_{k-2}\left(\re D_k^++s_k\tilde s_{k-2}\right)}\;.
\ee
There are two different roots at each $k=2,...\,, n-2$,
\be
\label{1root}
\re D_k^+ + s_k\tilde s_{k-2}  + \sigma_k = 0\;,
\ee
\be
\label{2root}
\re D_k^++s_k\tilde s_{k-2}   - \sigma_k \frac{s_k^2+\tilde s_{k-2}^2   - 2\sigma_k^{-1}s_k\tilde s_{k-2} }{s_k^2 + \tilde s_{k-2}^2 - 2 \sigma_k s_k \tilde s_{k-2}} = 0\;.
\ee
In what follows we choose just the first root \eqref{1root}, see our comments below Proposition \bref{equiv}. 


\bibliographystyle{JHEP}

\providecommand{\href}[2]{#2}\begingroup\raggedright
\addtolength{\baselineskip}{-3pt} \addtolength{\parskip}{-1pt}

\providecommand{\href}[2]{#2}\begingroup\raggedright\endgroup




\end{document}